\documentclass[11pt]{article}
\usepackage{amsmath,amssymb,amsthm,mathrsfs}
\usepackage{array}
\usepackage{graphicx,psfrag,epsf}
\usepackage{multirow}
\usepackage{multicol}
\usepackage{multirow}
\usepackage{booktabs}
\usepackage{natbib}
\usepackage{appendix}
\usepackage{float}
\usepackage[colorlinks,citecolor=blue,linkcolor=blue,urlcolor=blue]{hyperref}

\def\bbeta{\boldsymbol{\beta}}
\def\bX{{\mathbf X}}

\newtheorem{theorem}{Theorem}

\newtheorem{corollary}{Corollary}
\newtheorem{remark}{Remark}

\begin{document}

\title{Ensemble methods for testing a global null}

\author{Yaowu Liu\thanks{School of Statistics, Southwestern University of Finance and Economics, Chengdu, Sichuan, 611130, China.} \; Zhonghua Liu\thanks{Department of  Biostatistics, Columbia University, New York, NY, 10032, USA.}  \; and \; Xihong Lin\thanks{Department of Biostatistics, Harvard T.H. Chan School of Public Health, Boston, MA 02115, USA.}}

\date{}

\maketitle
\thispagestyle{empty}
\baselineskip=20pt

\begin{abstract}
Testing a global null is a canonical problem in statistics and has a wide range of applications. In view of the fact that no uniformly most powerful test exists, prior and/or domain knowledge are commonly used to focus on a certain class of alternatives to improve the testing power. However, it is generally challenging to develop tests that are particularly powerful against a certain class of alternatives. In this paper, motivated by the success of ensemble learning methods for prediction or classification, we propose an ensemble framework for testing that mimics the spirit of random forests to deal with the challenges. Our ensemble testing framework aggregates a collection of weak base tests to form a final ensemble test that maintains strong and robust power for global nulls. We apply the framework to four problems about global testing in different classes of alternatives arising from Whole Genome Sequencing (WGS) association studies. Specific ensemble tests are proposed for each of these problems, and their theoretical optimality is established in terms of Bahadur efficiency. Extensive simulations and an analysis of a real WGS dataset are conducted to demonstrate the type I error control and/or power gain of the proposed ensemble tests.
\end{abstract}

\noindent
 \textbf{Keywords:} Bahadur efficiency;  Cauchy P-value combination methods; Random weights; Robust test; Whole genome sequencing studies.

\pagestyle{plain}

\section{Introduction}\label{Sec:1}

\subsection{Global hypothesis testing}

Testing the global null that $H_0: \boldsymbol{\beta} = 0$ against $H_1: \boldsymbol{\beta} \neq 0$ is a canonical problem in statistics, where $\boldsymbol{\beta} = (\beta_1,\beta_2,\cdots,\beta_p)^T$ is a vector of parameters, e.g., $\beta_j$'s are the regression coefficients or the means of a multivariate normal distribution. It is well-known that uniform most powerful (UMP) test does not exist for global hypothesis testing. In fact, for testing a global null in multivariate normal models, any test will perform poorly under a wide range of alternatives of the composite hypothesis $H_1$~\citep[][Chapter 14.6]{lehmann2006testing}.

In view of these facts, a natural question is how one can improve the testing power in a specific application regarding global testing. A widely used approach is to leverage some prior and/or domain knowledge to focus on a certain class of alternatives that is of practical interests, and then develop tests that are particularly powerful against this class of alternatives.
For example, in whole-genome sequencing (WGS) association studies, one is interested in testing the association between a trait and a set of rare genetic variants, e.g., the variants in the promoter of a gene. The genetic variants in some sets, e.g., the putative loss of funtion (pLOF) variants of a gene, could be all protective or deleterious for a trait or disease~\citep{sham2014statistical}. This leads to the scenario of the same effect sign, i.e., the nonzero $\beta_j$'s are either all positive or negative under the alternative~\citep{li2008methods}.
Other scenarios of practical interest arising from WGS studies or other applications could be sparse signals~\citep{donoho2015special,barnett2017generalized}, different effect signs~\citep{lee2014rare}, weak signal strength~\citep{liu2020minimax}, and varied effect sizes of variables based on auxiliary information~\citep{wu2011rare}. Each of these scenarios corresponds to a class of alternatives that one would like to focus on.

The problem then becomes how to develop powerful tests in each of these
scenarios about the alternative hypothesis. However, it is generally challenging to use the traditional approaches, e.g., the likelihood ratio approach or the empirical Bayes approach~\citep[see, e.g.,][]{lin1997variance}, to deal with this problem, since the class of alternatives (e.g., the same effect sign) often corresponds to constraints on the parameter space. The constraints often make the effort of analytically deriving the optimal test statistics complicated or even mathematically intractable. In addition, even if the test statistic can be derived, it may follow a non-standard distribution under the null. This makes the p-value calculation computationally burdensome and dampens the practical utility of the test, especially for the large-scale multiple testing setting (see Section~\ref{Subsec:challenges:large:testing} for details about the computational challenges in this setting). These mathematical and computational challenges motivate us to develop powerful and computationally scalable tests.

\subsection{Motivation from ensemble learning}

In the area of prediction or classification, ensemble learning is a class of methods that have been proven to be powerful and efficient in a wide range of applications. Some well-known ensemble learning methods include random forests~\citep{breiman2001random}, bagging~\citep{breiman1996bagging}, and boosting~\citep{schapire1990strength,freund1995boosting}.
Ensemble learning methods first build a bank of simple base learners and aggregate them, e.g., by majority vote or averaging, to obtain a final learner. Individual base learners (e.g., tree) are often weak learners that may perform poorly for a prediction task, while the final leaner (e.g., random forest) combines the strengths of weak learners and greatly improves the predictive performance.

\cite{dietterich2002ensemble} pointed out that ensemble methods can partly overcome the problems encountered by the traditional approach of directly finding a single best learner. Roughly speaking, the task of prediction is to find a function that best approximates the underlying true function by searching through a functional space. Finding a single best learner (or classifier) directly encounters both mathematical and computational challenges, for example, the functional space may be too large for the amount of available training data, or learning algorithms cannot guarantee to obtain the best learner and may get stuck in a locally optimal solution~\citep{dietterich2002ensemble}.

We observe that these challenges are similar to those we encountered in the global testing problems, and the traditional approaches in testing can also be viewed as aiming to directly find a single  best test (in a certain sense). Given the success of ensemble methods in learning, we are interested in investigating whether an analog ensemble approach can  be developed to deal with the challenges in global testing. In this work, we
 develop powerful and scalable ensemble tests for the global testing problems mentioned above, with applications to large-scale rare variant association testing in WGS studies.

\subsection{The general framework of the proposed ensemble method for testing}\label{Subsec:framework}

As our ensemble method is motivated in the spirit of random forest, we first provide a brief review of the random forest algorithm and point out its key steps.
Let $T_{\text{tree}}(\mathbf{x}, \mathcal {D},\Theta_i)$ denote the $i$-th random tree, where $\mathbf{x}$ represents the features, $\mathcal {D}$ denotes the data, and $\Theta_i$ is the random component in the process of building the $i$-th tree, including drawing a bootstrap sample and randomly selecting variables in node splitting. The $\Theta_i$'s are independently and identically distributed (i.i.d.). Suppose that $B$ trees are grown.
For regression, the random forest takes the average of the $B$ trees and is given by
\[
  \hat{f}_{\text{rf}} (\mathbf{x}) = \frac{1}{B}\sum_{i=1}^B T_{\text{tree}}(\mathbf{x}, \Theta_i,\mathcal {D}).
\]
For the details of random forest, we refer readers to~\cite{breiman2001random},~\cite{hastie2009elements} and the references therein.

 Developing an ensemble  procedure for hypothesis testing by mimicking ensemble learning methods  is conceptually simple and straightforward, i.e., construct some base tests and then combine them to obtain a final ensemble test. But two key questions need to be addressed for ensemble testing. The first question is how to construct base tests, and the second one is how to combine the base tests.

\textit{Construction of base tests.} To address the first question, we mimic the way of constructing a base learner in random forests. Let $T_{\text{stat}}(\mathcal {D},\Theta_i)$ denote the $i$-th base test statistic, where $\mathcal {D}$ denotes the data, and $\Theta_i$ is a certain random procedure (e.g., random weighting). The $\Theta_i$'s are also required to be i.i.d. and are independent of the data $\mathcal {D}$. In random forests, the random component $\Theta_i$ (i.e., bootstrapping and random selection of variables) is crucial to its success in practice. Similarly, the random component $\Theta_i$ is the key part in our ensemble testing framework. First and foremost, it is what we use to characterize the scenario about the alternatives (e.g., the same effect sign), and to circumvent the challenges faced by the traditional approaches of finding a single best test directly. Second, it helps to produce a diverse set of individual base tests such that the ensemble of them would be beneficial. The choices of $\Theta_i$ and the test statistic $T_{\text{stat}}$ would depend on the specific problem. In this paper, we study several concrete problems about global testing in different scenarios of the alternatives.
For each of these problems, we propose to use random weighting or random selection of variables as $\Theta_i$ and use existing popular tests as the base tests $T_{\text{stat}}$. The details will be discussed in the following sections.

\textit{Combination of base tests.} To address the second question, as the p-values (or test statistics) of the base tests are calculated based on the same data $\mathcal {D}$ and therefore are correlated, we aggregate them via the ACAT (Aggregated Cauchy Association Test) method~\citep{liu2019acat,liu2020cauchy}, which is particularly advantageous for combining correlated p-values.
Specifically,
let $p(\mathcal {D}, \Theta_i)$ denote the p-value corresponding to the base test statistic $T_{\text{stat}}(\mathcal {D},\Theta_i)$. The ensemble test statistic is
\begin{equation}\label{eq:ETstatistic}
  T (\mathcal {D}) = \frac{1}{B}\sum_{i=1}^{B} h_{\text{acat}} \{ p(\mathcal {D}, \Theta_i) \},
\end{equation}
where $h_{\text{acat}}(p)= \tan\{(0.5-p)\pi\}$ is a function in ACAT that transforms the p-value to a standard Cauchy variable under the null. A notable feature of ACAT is that the combined p-value can be approximated by a Cauchy distribution without the need to account for the correlations among the individual p-values. Hence, for the ensemble test statistic $T (\mathcal {D})$, we can simply calculate its p-value as
\begin{equation}\label{eq:pval:approx}
  p_{\text{en}} = 1/2 - \{\arctan T (\mathcal {D})\}/\pi.
\end{equation}
 In addition, ACAT is a powerful aggregation approach for combining correlated base tests. Many base tests could have a low power against the alternative of the unknown true signal. Since ACAT mainly uses several smallest p-values of the base tests to represent the overall significance~\citep{liu2019acat}, the underpowered base tests would only have a limited impact on the power of the final ensemble test.
More details about ACAT are provided in Section~\ref{Secsub:ACAT} and the choice of the number of base tests $B$ will be discussed in Section~\ref{Secsub:B}.

The proposed ensemble framework provides a flexible way to construct powerful tests targeted at a certain class of alternatives.
In this paper, we will apply this ensemble framework to four global testing problems in different scenarios about the class of alternatives,
i.e., testing under the same effect sign, testing under different effect signs, testing against sparse alternatives, and testing with auxiliary information about relative effect sizes of variables. In particular, we will provide a detailed study for the problem of testing under the same effect sign (which is our initial motivation to explore ensemble methods for testing), and use it to demonstrate the rationale and many other aspects of the proposed ensemble testing framework. Specific ensemble tests will be proposed for each of the four problems.
For all the proposed ensemble tests, we will show that their type I errors are controlled when the significance level goes to 0, and investigate their theoretical power in terms of Bahadur efficiency~\citep{bahadur1960stochastic}.

\subsection{The motivating application}\label{Subsec:motivating:application}
The global testing problems considered here could arise from a wide range of applications, such as copy number analysis~\citep{jeng2010optimal}, microbiome-profiling studies~\citep{zhao2015testing}, and transcriptome-wide association studies~\citep{feng2021leveraging}. Our work
is motivated from rare variant association tests (RVATs) in Whole-Genome Sequencing (WGS) studies, where variant set analysis~\citep{lee2014rare} is commonly conducted to investigate whether a set  of genetic variants (Single Nucleotide Variants (SNVs)) is associated with a disease or phenotype. The variant sets are typically defined based on some biological knowledge. For instance, one could group  the nonsynonymous variants or variants in the enhancer of a gene, say, LDLR (low-density lipoprotein receptor), into a set, and test their effects on low-density lipoprotein(LDL) cholesterol~\citep{li2020dynamic}.  Tens of thousands to millions of variant sets are scanned across the genome in WGS studies, with the association test for each variant set corresponding to a global testing problem.  To improve the power of detection, various prior knowledge~\citep{sham2014statistical} and/or auxiliary information~\citep{wu2011rare,li2020dynamic}  have been used to focus on particular classes of alternatives that are of practical interests. This leads to a pressing need to develop powerful tests in these different scenarios of the classes of alternatives.

\subsection{Computational challenges in large-scale multiple testing}\label{Subsec:challenges:large:testing}
In WGS studies, a global test needs to be applied to millions of variant sets across the genome. This application is an example of global testing in the context of large-scale multiple testing, which poses considerable computational challenges in test development.
First, as a test needs to be applied a massive number of times, fast computation of the test is required for the test to be practically useful.
Second, since the significance threshold would be exceedingly small to account for multiple testing (e.g., the genome-wide significance threshold is around  $\alpha=10^{-8}$), accurate p-value calculation is needed when the p-value is tiny. While resampling methods (e.g., permutation) are generally applicable for calculating p-values, they are computationally burdensome for evaluating extremely small p-values and therefore may not
be feasible for large-scale testing. Also, permutation cannot handle related samples, which are common in WGS studies. These computational challenges have already motivated substantial efforts to speed up the computation of some existing popular tests~\citep[see, e.g.,][]{wu2011rare, barnett2017generalized,zhang2022generalized}. We note that the computational challenges described above in large-scale multiple testing are quite different from those in the high-dimensional setting, and we are primarily interested in the large-scale multiple testing setting in this work, motivated by WGS association studies.

In summary, we aim to develop testing methods that not only improve the power over the existing methods, but also are computationally scalable for the large-scale multiple testing setting, and also have good theoretical properties. In other words, we would like the proposed tests to achieve the three goals simultaneously.

\subsection{Related work}

A variety of tests have been proposed for the testing problems considered here, e.g., the Burden test~\citep{li2008methods,madsen2009groupwise} for testing under the same effect sign, the Sequence Kernel Association Test (SKAT)~\citep{wu2011rare} and Minimax Optimal Ridge-Type Set Test (MORST)~\citep{liu2020minimax} for testing under different effect signs, and the Higher Criticism~\citep{donoho2004higher} and Berk-Jones tests~\citep{berk1979goodness} for testing against sparse alternatives. While these existing tests are useful and advantageous in various aspects, each of them has their own limitations that will be discussed in details in the subsequent sections. Overall, none of the existing tests (e.g., the Burden test) guarantees to be the best test in their respective scenarios (e.g., the same effect sign),
and hence it is desirable to improve their powers. In addition, these existing tests are developed on a case-by-case basis, and
there lacks a general and unified approach for developing powerful tests targeted at a certain class of alternatives.

Omnibus testing is a popular practical method to derive robust tests through test combination~\citep[e.g.,][]{lee2012optimal,zhu2015meta,zhao2015testing,mccaw2020operating,feng2021leveraging}, and can also be viewed as an ensemble approach. Our ensemble method is different from the omnibus approach in two aspects. First of all, the goals of the two ensemble approaches are distinct.
The omnibus approach aims to derive tests robust to different scenarios. For example, \cite{lee2012optimal} combined a test advantageous under the same effect sign (i.e., the Burden test) and a test advantageous under different effect signs (i.e., SKAT) to derive an omnibus test that has robust power to the two scenarios.
But the omnibus approach does not address the problem of developing powerful tests in each of these scenarios (e.g., a powerful test under the same effect sign), which is the goal of our ensemble testing methods. Second, the two ensemble approaches also differ in the tests being combined. The omnibus test often combines several heterogenous tests (i.e., different types of tests), which is parallel to ensemble learners that combine heterogenous base learners (e.g., combining k nearest neighbor, naive Bayes and support vector machine) in ensemble learning. In contrast, our ensemble approach combines a large number of homogenous tests (i.e., tests that are of the same type), and is parallel to ensemble learners (e.g., random forest) that combine homogenous base learners.

Testing methods based on random projection have been proposed in the literature~\citep[see, e.g.,][]{srivastava2016raptt}. The random weighting method used in the construction of base tests can be viewed as a random projection of a vector of test statistics to a scalar.  We note that there is a fundamental difference between random projection and our random weighting method. The key rationale of random projection arises from the Johnson-Lindenstrauss (JL) lemma~\citep{johnson1984extensions}: a set of points in the $p$-dimensional space can be randomly projected onto a $k$-dimensional subspace such that the pairwise Euclidian distances between the points are approximately preserved with a high probability~\citep{vempala2005random}. As can be seen from the JL lemma, while $k$ can be much smaller than $p$,  there is a lower bound for $k$, which requires that the dimension $k$ of the subspace should be suitably large in order to preserve the Euclidian distances. In particular, $k$ needs to be much larger than 1. In contrast, our random weighting method projects a vector of test statistics onto the $1$-dimensional subspace. Therefore, although can be viewed as random projection, the rationale of random weighting here is completely different from that of random projection.

~\cite{escanciano2006consistent} and~\cite{sun2021projection} proposed methods to randomly project vectors to scalars like ours.  However, their methods differ from our ensemble approach in multiple folds. First, their goal is to perform model checking for partially linear or parametric models while we focus on testing the globe null $H_0:\bbeta=0$, and their test statistics are different. Second, they aggregate their random weight projected tests by uniformly integrating out the weights on the unit sphere, which is equivalent to calculating the averages of replicates of the random weight based statistics across all possible directions. This aggregation approach will be  subject to considerable power loss for our problem, and is also computationally intensive as it cannot be calculated analytically. In contrast, we aggregate random-weighted base tests using ACAT, which has attractive power benefits when aggregating many underpowered correlated base tests (see Section~\ref{Secsub:ACAT}), and is computationally very fast.  Third, the rationale of random weights in their methods is based on an equivalent representation of the null hypothesis that is specific to partially linear or parametric models. Hence,  the goal, test statistic, aggregation strategy, and rationale of their methods are all different from ours.

\subsection{Outline}

Section~\ref{Sec:Burden} presents the ensemble Burden test for testing under the same effect sign and its theoretical properties, and discusses many other aspects of the general ensemble framework.
Sections~\ref{Sec:aux:info}--\ref{Sec:chisq} present the specific ensemble tests for the other three global testing problems and their theoretical properties.
Section~\ref{Sec:simu} demonstrates the effectiveness of our proposed ensemble tests through simulations designed to mimic realistic WGS data, and Section~\ref{Sec:realdata} illustrates the power of our ensemble tests when applied to a real WGS dataset from the Atherosclerosis Risk in Communities(ARIC) study.
Concluding remarks and possible further research directions are given in Section~\ref{Sec:diss}.
All the technical proofs and additional simulation results are relegated to the Supplementary Materials.

\section{The Ensemble Burden test}\label{Sec:Burden}

\subsection{Bahadur efficiency}\label{Secsub:Bahadur}

As we will study Bahadur efficiency~\citep{bahadur1960stochastic} of the proposed ensemble tests in Sections~\ref{Sec:Burden}-\ref{Sec:chisq}, we first provide an overview of it. Let $N_i$ be the smallest sample sizes required by the test $T_i$ to achieve a power equal to $\gamma$ under a given alternative $\boldsymbol{\beta}$ at the significance level $\alpha$, where $i=1,2$. The efficiency of the test $T_1$ relative to the test $T_2$ is the ratio of the smallest sample sizes, i.e., $\text{Eff}(T_1,T_2) = N_2(\alpha,\gamma,\boldsymbol{\beta})/N_1(\alpha,\gamma,\boldsymbol{\beta})$. If $\text{Eff}(T_1,T_2)>1$. This indicates the test $T_1$ requires a smaller sample size and therefore is more efficient than $T_2$. Except for a few simple cases, calculating the efficiency is generally difficult or even mathematically intractable. Therefore, several extensions have been proposed.
Bahadur efficiency concerns the efficiency in the limiting situation where $\alpha$ goes to 0, i.e., $\text{Eff}_b(T_1,T_2) = \lim_{\alpha\rightarrow 0}N_2(\alpha,\gamma,\boldsymbol{\beta})/N_1(\alpha,\gamma,\boldsymbol{\beta})$. When $\alpha$ goes to 0, the power of any test would vanish to 0 if the sample size does not go to infinity. Hence, Bahadur efficiency essentially compares how fast the sample size should grow in order to let the two tests have the same power $\gamma$. Typically, Bahadur efficiency does not depend on $\gamma$ and also may not depend on $\boldsymbol{\beta}$. For more discussions about Bahadur efficiency, we refer readers to~\cite{dasgupta2008asymptotic},~\cite{van2000asymptotic} and the references therein.

\subsection{The model}\label{Secsub:basic:Burden}
Testing under the same effect sign is a common problem arising from the situations where the effects of variables are expected to be all positive or negative. As introduced in Section~\ref{Sec:1}, one such an example is from the variant set analysis in WGS studies, where the Burden test~\citep{li2008methods,madsen2009groupwise,price2010pooled} has emerged as a popular tool in this area, especially for testing the effects of the set of putative loss-of-functional (pLOF) variants of a gene on a phenotype \citep{li2020dynamic}. Another application where Burden test is widely used is copy number analysis~\citep{jeng2010optimal,zhang2010detecting}, where the effects of copy numbers in a local region are often in the same direction as copy number gains or losses often happen together locally.

To fix the idea and simplify the notation, we first consider a simplified linear regression model with a known error variance:
\begin{equation}
   \mathbf{Y}=\mathbf{X}\boldsymbol{\beta}+\boldsymbol{\varepsilon},
\label{eq:model}
\end{equation}
where $\mathbf{Y} \in \mathbb{R}^n$ is a vector of responses, $\mathbf{X} \in \mathbb{R}^{n\times p}$ is a fixed design matrix with centered columns, $\boldsymbol{\beta}\in \mathbb{R}^p$ is a vector of coefficients, and $\boldsymbol{\varepsilon}\in \mathbb{R}^n$ is a vector of i.i.d. error terms following a standard normal distribution $N(0,1)$. We are interested in testing
\begin{equation}\label{eq:test:problem}
  H_0: \boldsymbol{\beta}= 0  \quad\quad \text{against}  \quad \quad H_1: \boldsymbol{\beta} \neq 0 \;\text{and all the nonzero}\; \beta_j\text{'s have the same sign},
\end{equation}
where $\beta_j$'s are the components of $\boldsymbol{\beta}$. The constraint in $H_1$ reflects our prior (or domain) knowledge that all the variables have the same effect sign, and incorporation of the prior knowledge could improve the power.

Let $\mathbf{S} = \frac{1}{\sqrt{n}} \mathbf{X}^T\mathbf{Y}$ denote marginal score test statistics for the regression coefficients $\beta_j$'s. We have
\begin{equation}\label{eq:simple:model}
   \mathbf{S} \sim N_p (\sqrt{n}\boldsymbol{\Sigma}\boldsymbol{\beta},\boldsymbol{\Sigma}),
\end{equation}
where $\boldsymbol{\Sigma} = \frac{1}{n}\mathbf{X}^T\mathbf{X}$. We note that testing the global null  in linear models with an unknown error variance, in generalized linear models for continuous and discrete outcomes, and a wide range of problems can (asymptotically) boil down to model~\eqref{eq:simple:model}\citep[see, e.g.,][]{lee2014rare,liu2020minimax}. Hence, we will directly work with model~\eqref{eq:simple:model} hereafter.

\subsection{The Burden test and its limitation}\label{Secsub:Burden}
Suppose that we only know the variables have the same effect sign but there is no additional knowledge about their effect sizes $|\beta_j|$'s. As all the $\beta_j$'s have the same sign, a natural choice of a test statistic is to simply sum up the components of $\mathbf{S}$. This results in the Burden test statistic \citep{li2008methods,lee2014rare}, i.e., $ T_{\text{Burden}} = \mathbf{1}_p^T\mathbf{S}$, where $\mathbf{1}_p$ denotes a $p$-dimensional vector of ones. In addition, since the common sign of $\beta_j$'s is typically unknown, a two-sided test based on $|T_{\text{Burden}}|$ is used and is referred to as the Burden test. Under the alternatives where all the $\beta_j$'s are equal, it can be easily seen that the Burden test is a uniformly most powerful unbiased (UMPU) test.

For any  true $\boldsymbol{\beta}$, we can interpret $||\boldsymbol{\beta}||$ as the true signal strength and $\mathbf{w}_\beta = \boldsymbol{\beta}/||\boldsymbol{\beta}||$ as the signal direction.
Under alternatives with the same signal direction $\pm \mathbf{w}_\beta$ but different signal strengthes,  the two-sided test based on $T_\beta = \mathbf{w}_\beta^T\mathbf{S}$ is the UMPU test.  Therefore, only the signal direction $\mathbf{w}_\beta$ needs to be considered. In practice, as $\mathbf{w}_\beta$ is unknown, this oracle test cannot be constructed.
Let $\mathcal{S}_p^{+} = \{(w_1,w_2,\cdots,w_p): \sum_{j=1}^pw_j^2=1, w_j\geq 0,\forall j=1,2,\cdots,p\}$ denote the $p$-dimensional unit sphere in the nonnegative orthant. The unit sphere in the nonpositive orthant $\mathcal{S}_p^{-}$ can be similarly defined by replacing $w_j \geq 0$ in $\mathcal{S}_p^{+}$ with $w_j \leq 0$. Then, $\mathcal{S}_p^{+}\cup \mathcal{S}_p^{-}$ represents the collection of all the possible directions of $\boldsymbol{\beta}$ when $\beta_j$'s have the same sign. If we have no knowledge about the relative magnitudes of $\beta_j$'s, this is equivalent to saying that there is no knowledge about the signal direction $\mathbf{w}_\beta$. The Burden test reflects this scenario by treating all the marginal score statistics equally, i.e., $ T_{\text{Burden}} = \mathbf{1}_p^T\mathbf{S}$. While simple and intuitive, the Burden test actually could lose considerable power  under certain dependency structures of $\mathbf{X}$,
 even when the effects have the same sign. We next provide theoretical analyses and examples to gain some insights about the reason.

For any unit vector $\mathbf{w} \in \mathbb{R}^p$, we have
\begin{equation}\label{Eq:linear:normal:mean}
  \frac{\mathbf{w}^T\mathbf{S}}{\sqrt{\mathbf{w}^T \boldsymbol{\Sigma} \mathbf{w}}} \sim N\left(\sqrt{n}||\boldsymbol{\beta}|| \cdot \frac{\mathbf{w}^T\boldsymbol{\Sigma}\mathbf{w}_\beta}{\sqrt{\mathbf{w}^T\boldsymbol{\Sigma}\mathbf{w}}},1\right).
\end{equation}
Hence, the power of any linear test $\mathbf{w}^T\mathbf{S}$ is solely determined by the absolute value of the mean of the normal distribution in~\eqref{Eq:linear:normal:mean}.
Furthermore, $\boldsymbol{\Sigma}$ can be decomposed as $\boldsymbol{\Sigma} = \mathbf{U}\boldsymbol{\Lambda}\mathbf{U}^T$, where $\boldsymbol{\Lambda}$ is a diagonal matrix of eigenvalues and the columns of $\mathbf{U}$ are the corresponding eigenvectors. Let $\bar{\mathbf{w}} = \boldsymbol{\Lambda}^{1/2}\mathbf{U}^T\mathbf{w}$ and $\bar{\mathbf{w}}_\beta = \boldsymbol{\Lambda}^{1/2}\mathbf{U}^T\mathbf{w}_\beta$ be the transformed vectors of $\mathbf{w}$ and $\mathbf{w}_\beta$ based on $\boldsymbol{\Sigma}$, respectively.
Then, the normal mean in~\eqref{Eq:linear:normal:mean} can be further written as $\sqrt{n}||\boldsymbol{\beta}|| \cdot ||\bar{\mathbf{w}}_\beta|| \cdot \cos \theta(\bar{\mathbf{w}},\bar{\mathbf{w}}_\beta)$, where $\theta(\bar{\mathbf{w}},\bar{\mathbf{w}}_\beta)$ denotes the angle between $\bar{\mathbf{w}}$ and $\bar{\mathbf{w}}_\beta$. Note that only $\theta(\bar{\mathbf{w}},\bar{\mathbf{w}}_\beta)$ in the normal mean is affected by our choice of $\mathbf{w}$. Hence, the efficiency of a linear test $T_w = \mathbf{w}^T\mathbf{S}$ relative to the optimal oracle test $T_\beta = \mathbf{w}_\beta^T\mathbf{S}$ only depends on the angle $\theta(\bar{\mathbf{w}},\bar{\mathbf{w}}_\beta)$, which is the angle between our selected direction in $T_w$ and the unknown true signal direction after $\boldsymbol{\Sigma}$-based transformation. Specifically, it can be easily seen that $\text{Eff}(T_w,T_\beta) = \cos^2 \theta(\bar{\mathbf{w}},\bar{\mathbf{w}}_\beta)$.

Let $\mathbf{w}_{1_p} = \mathbf{1}_p/\sqrt{p}$, which is the direction the Burden test uses. To see the limitations of the Burden test, consider the case where $\boldsymbol{\Sigma} = \{\sigma_{ij}\}$ is an exchangeable correlation matrix, i.e., $\sigma_{ii}=1$ and $\sigma_{ij}=\rho$ for $1\leq i \neq j \leq p$. In this case, $\mathbf{w}_{1_p}$ is the first eigenvector of $\boldsymbol{\Sigma}$ and its corresponding eigenvalue is $\lambda_1=1+(p-1)\rho$. The other eigenvalues of $\boldsymbol{\Sigma}$ are the same, i.e., $\lambda_2=\lambda_3=\cdots=\lambda_p = 1-\rho$. Since all the eigenvalues must be non-negative, the valid values for $\rho$ are $-1/(p-1)\leq\rho\leq1$. Next, we consider the efficiency of Burden test relative to the optimal test $T_\beta$. Using the simple fact that $\boldsymbol{\Sigma}\mathbf{w}_{1_p} = \lambda_1\mathbf{w}_{1_p}$, we have
\begin{equation}\label{eq:burdenExchange}
\cos \theta(\bar{\mathbf{w}}_{1_p},\bar{\mathbf{w}}_\beta) = \frac{\lambda_1 \cos \theta(\mathbf{w}_{1_p},\mathbf{w}_\beta)}{||\bar{\mathbf{w}}_\beta||} = \frac{\cos \theta(\mathbf{w}_{1_p},\mathbf{w}_\beta)}{\sqrt{\cos^2 \theta(\mathbf{w}_{1_p},\mathbf{w}_\beta)+(\lambda_2/\lambda_1) \sum_{j=2}^p (\mathbf{U}_j^T\mathbf{w}_\beta)^2} },
\end{equation}
where $\theta(\mathbf{w}_{1_p},\mathbf{w}_\beta)$ denotes the angle between $\mathbf{w}_{1_p}$ and $\mathbf{w}_\beta$, and $\mathbf{U}_j$'s are the eigenvectors of $\boldsymbol{\Sigma}$. Note that only $\lambda_2/\lambda_1 = (1-\rho)/[1+(p-1)\rho]$ in~\eqref{eq:burdenExchange} depends on $\rho$ and recall that $|\cos \theta(\bar{\mathbf{w}}_{1_p},\bar{\mathbf{w}}_\beta)|$ represents the efficiency of the Burden test. When $\rho$ is positive and approaches 1 (i.e., $\lambda_2$ is close to 0), the Burden test has high efficiency no matter what the signal direction $\mathbf{w}_\beta$ is. However,
for any signal direction $\mathbf{w}_\beta \neq \mathbf{w}_{1_p}$, as long as $\rho$ is sufficiently close to its lower bound $-1/(p-1)$ (i.e., $\lambda_1$ is close to 0), the efficiency of the Burden test would be close to 0. In other words, when $\boldsymbol{\Sigma}$ has a negative exchangeable correlation, the Burden test would have low efficiency at majority of signal directions in $\mathcal{S}_p^{+}$ and therefore is not appropriate in this situation.

\begin{figure}[!h]
  \centering
  \includegraphics[scale=0.50]{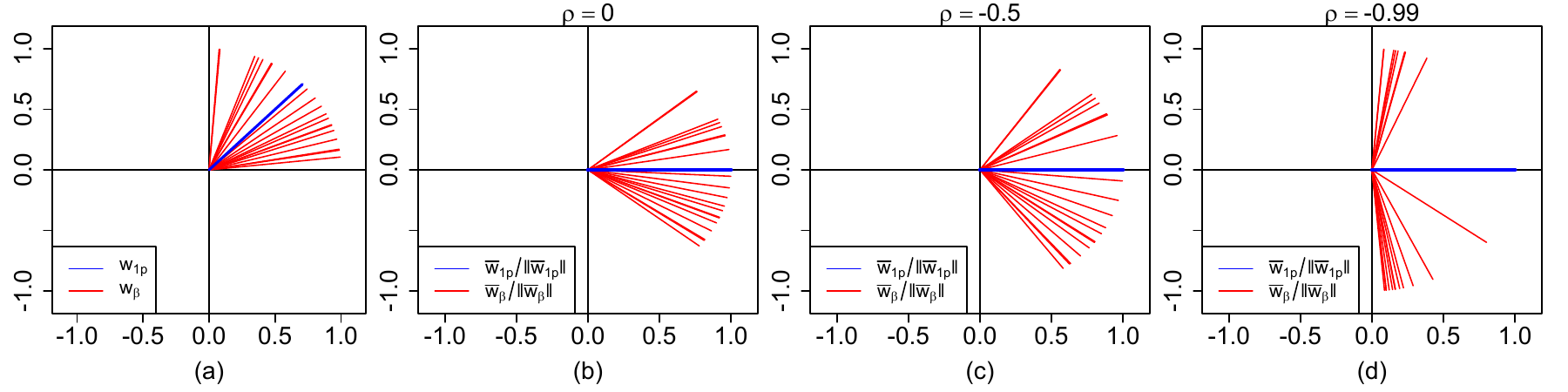}
  \caption{A 2-dimensional illustration of the limitation of the Burden test under exchangeable correlations of $\mathbf{X}$. The red lines show 20 true signal directions $\mathbf{w}_\beta$'s randomly drawn from $\mathcal{S}_p^+$. The blue line is the Burden test direction $\mathbf{w}_{1_p}$. Figure (a) displays the signal directions and Burden test direction before transformation. Figures (b), (c) and (d) show the transformed directions (i.e., $\bar{\mathbf{w}}_{1_p}$ and $\bar{\mathbf{w}}_\beta$) under the exchangeable correlation of $\bX$ with $\rho = 0, -0.5, -0.99$, respectively. }
  \label{Fig:Burden_limitation}
\end{figure}

Figure~\ref{Fig:Burden_limitation} provides a 2-dimensional illustration of this analysis. Figure~\ref{Fig:Burden_limitation}(a) shows 20 signal directions $\mathbf{w}_\beta$'s randomly drawn from $\mathcal{S}_p^+$ and the Burden direction $\mathbf{w}_{1_p}$ before transformation. Figure~\ref{Fig:Burden_limitation}(b), (c) and (d) display the transformed directions under exchangeable correlations with $\rho = 0, -0.5, -0.99$, respectively. Specifically, $\bar{\mathbf{w}}_{1_p} = \boldsymbol{\Lambda}^{1/2}\mathbf{U}^T\mathbf{w}_{1_p}$ and $\bar{\mathbf{w}}_\beta = \boldsymbol{\Lambda}^{1/2}\mathbf{U}^T\mathbf{w}_\beta$, where $\mathbf{U} = \begin{pmatrix} 1/\sqrt{2}& -1/\sqrt{2} \\ 1/\sqrt{2} & 1/\sqrt{2} \end{pmatrix}$ and $\boldsymbol{\Lambda} = \begin{pmatrix} 1 + \rho & 0 \\ 0 & 1 - \rho \end{pmatrix}$.
As $\rho$ decreases, we can see the angles between $\bar{\mathbf{w}}_\beta$ and $\bar{\mathbf{w}}_{1_p}$ become larger, which implies a smaller efficiency of the Burden test. When $\rho$ is close to $-1$, the Burden test has low efficiency for most of signal directions in $\mathcal{S}_p^+$.

The analysis above indicates that the Burden test is not always an appropriate choice and could even be almost powerless for testing under the same effect sign.
Geometrically, the direction the Burden test uses, i.e., $\mathbf{w}_{1_p}$, is in the ``center" of $\mathcal{S}_p^{+}$. As the true signal direction $\mathbf{w}_\beta \in \mathcal{S}_p^{+}$ is unknown, $\mathbf{w}_{1_p}$ seems to be a natural choice to achieve robust power across different signal directions. This makes the Burden test intuitively appealing at the first glance.
  However, as shown in the theoretical analysis, the power of linear tests is not determined by the angle between the signal direction and our selected direction (i.e., $\theta(\mathbf{w},\mathbf{w}_\beta)$), but the angle after $\boldsymbol{\Sigma}$-based transformation (i.e., $\theta(\bar{\mathbf{w}},\bar{\mathbf{w}}_\beta)$). After transformation, $\bar{\mathbf{w}}_{1_p}$ may not be the geometrical ``center" any more. In fact, in the extreme case where $\boldsymbol{\Sigma}$ has a negative exchangeable correlation, the angle $\theta(\bar{\mathbf{w}}_{1_p},\bar{\mathbf{w}}_\beta)$ is large for most signal directions. This explains why the Burden test could perform poorly under certain $\boldsymbol{\Sigma}$.

\subsection{The Ensemble Burden test}\label{Secsub:ENBurden}
In view of the limitation of the Burden test, one would ask how to develop a more powerful test for the composite alternative hypothesis \eqref{eq:test:problem}. One approach is to consider the likelihood ratio test (LRT).
Due to the constraint posed by the same effect sign under $H_1$,  the LRT statistic does not have a closed form. Also, Wilk's theorem is not applicable since the full parameter space is a closed set~\citep[][Chapter 16.1, p.p. 228]{van2000asymptotic}, and re-sampling methods might be needed for calculating the p-value of LRT. These makes the likelihood ratio approach computationally burdensome for the large-scale multiple testing setting in WGS studies mentioned in Section~\ref{Subsec:challenges:large:testing}, which we are primarily interested in. More details about the computational challenges of the LRT are provided in Section 9 in the Supplementary Materials. In this work,
we use the ensemble framework described in Section~\ref{Subsec:framework} to construct a powerful and computationally scalable test for testing under the same effect sign (\ref{eq:test:problem}).

The general idea is as follows. Since no knowledge about the true signal diretion $\mathbf{w}_\beta$  means that all the directions in $\mathcal{S}_p^{+}\cup \mathcal{S}_p^{-}$ are equally possible, we first randomly draw a collection of directions from $\mathcal{S}_p^{+}\cup \mathcal{S}_p^{-}$ with equal probability (this step corresponds to the random component $\Theta_i$ in the framework). Then, for each direction, we obtain its corresponding linear test and use it as the base test (this corresponds to the base test statistic $T_{\text{stat}}$ in the framework). In fact, we only need to consider the directions in $\mathcal{S}_p^{+}$ as two-sided linear tests are employed. Finally, we combine the p-values of the base tests via ACAT to obtain the final ensemble burden test.

Specifically, let $\mathbf{w}_i$'s be i.i.d. random vectors following a uniform distribution on $\mathcal{S}_p^{+}$, where $i=1,2,\cdots,B$. The multivariate Gaussian distribution can be used to simulate such random vectors. Specifically, we first generate independent Gaussian vectors $\boldsymbol{\xi}_i \sim N_p(0,\mathbf{I})$, and then let $\mathbf{w}_i  = (\frac{|\xi_{i1}|}{||\boldsymbol{\xi}_i||},\frac{|\xi_{i2}|}{||\boldsymbol{\xi}_i||},\cdots,\frac{|\xi_{ip}|}{||\boldsymbol{\xi}_i||})^T$. It can be easily shown $\mathbf{w}_i$ is uniformly distributed on $\mathcal{S}_p^{+}$.
 Let
\begin{equation*}
  T_i = \frac{\mathbf{w}_i^T\mathbf{S}}{\sqrt{\mathbf{w}_i^T \boldsymbol{\Sigma} \mathbf{w}_i}},
\end{equation*}
be the $i$-th test statistic. A two-sided test based on $|T_i|$ is the UMPU test if the signal direction $\mathbf{w}_\beta=\pm \mathbf{w}_i$ and serves as the base test in the ensemble test.
The ensemble Burden test statistic $T_{\text{EN-Burden}}$ then aggregates the p-values of $T_i$'s through the ACAT method and is given by
\[
    T_{\text{EN-Burden}}=\frac{1}{B}\sum_{i=1}^B \tan\{ [2\Phi(|T_i|)-3/2]\pi \}.
\]
Finally, we approximate the p-value of the ensemble Burden test by~\eqref{eq:pval:approx} based on the standard Cauchy distribution.

 Since $\mathbf{w}_i$'s are randomly and independently drawn from $\mathcal{S}_p^{+}$, the random component in the ensemble Burden test characterizes the feature of the class of alternatives in the scenario of the same effect sign. The idea behind the proposed random component (i.e., random weighting) is to explore a variety of linear tests with different directions to be robust to the unknown true signal direction in $\mathcal{S}_p^{+}$. Additional explanations about the improvement of the ensemble test will be provided in Section~\ref{Secsub:explanation}.

\begin{figure}[!h]
  \centering
  \includegraphics[scale=0.65]{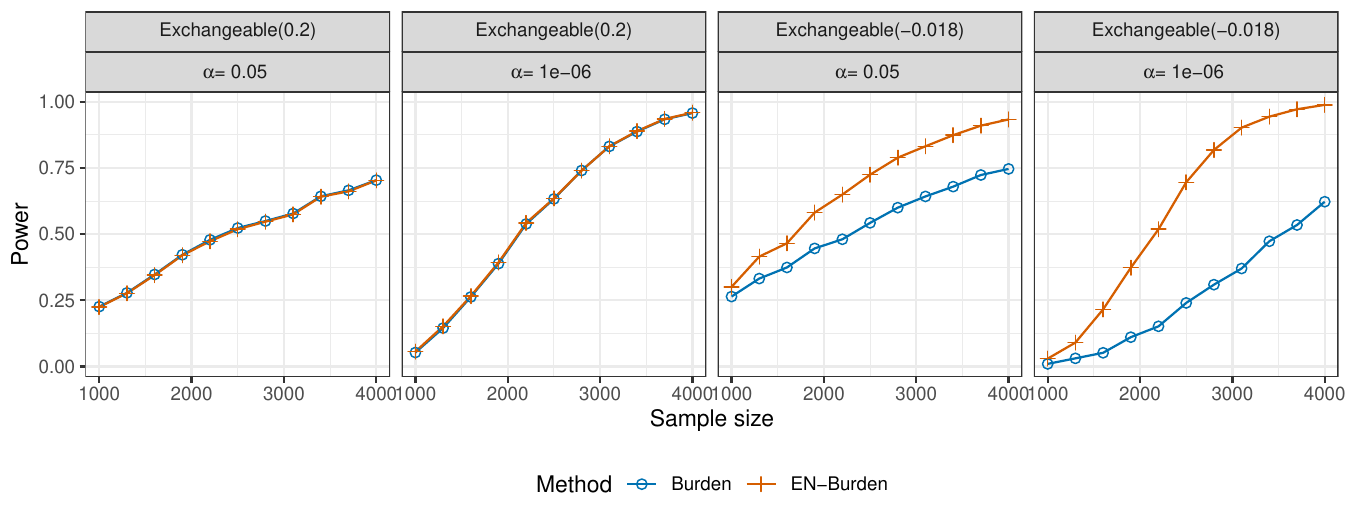}
  \caption{The power of Burden and ensemble Burden tests in a simulation example. Here, the model is $\mathbf{S} \sim N_p (\sqrt{n}\boldsymbol{\Sigma}\boldsymbol{\beta},\boldsymbol{\Sigma})$ with $\boldsymbol{\Sigma}$ being an exchangeable correlation matrix, where the exchangeable correlation $\rho = 0.2$ on the left two panels or $\rho = -0.018$ on the right two panels. In each replication, the signal direction $\mathbf{w}_\beta$ is independently and randomly drawn from $\mathcal{S}_p^{+}$. The signal strength $||\boldsymbol{\beta}||$ is set to be 0.015, 0.15, 0.04 and 0.3 for the four panels from left to right, respectively. The y-axis is the power under 2000 replications and the x-axis is the sample size $n$.
  The dimension of $\bX$ is $p=50$. The number of base tests $B=1000$.
  Significance levels $\alpha=0.05$ and $\alpha=10^{-6}$ are considered. When $\alpha=0.05$, the critical values of the ensemble Burden test are obtained through simulations to make a fair comparison. }
  \label{PowerExampleplots}
\end{figure}

 Now, we go back to the exchangeable correlation example described in Section~\ref{Secsub:Burden} and compare the power of the Burden and ensemble Burden tests. Figure~\ref{PowerExampleplots} shows the results of a simulation where the exchangeable correlation $\rho$ is $0.2$ or $-0.018$ and the signal directions are randomly drawn from $\mathcal{S}_p^{+}$ . When $\rho=0.2$ is positive, the ensemble Burden test is as powerful as the Burden test. When $\rho = -0.018$ is negative, the Burden test loses substantial power compared to the ensemble Burden test. The results are consistent with our theoretical analysis in Section~\ref{Secsub:Burden}. As the ensemble Burden test explores many different directions by combing the strengthes of the base tests, it is robust to the signal direction and hence improves the overall power.
In summary, these results demonstrate that the ensemble Burden test provides a more effective and robust way to reflect the scenario of the same effect sign than the Burden test.

We next investigate the theoretical power of the ensemble Burden test in terms of Bahadur efficiency, whose basic concept has been introduced in Section~\ref{Secsub:Bahadur}. We first consider the type I error and the theorem below indicates that the null distribution of $T_{\text{EN-Burden}}$ is a standard Cauchy distribution in the tail.
\begin{theorem}\label{Thm:typeI:burden}
For any fixed $B\geq 1$ and any fixed realizations of $\mathbf{w}_i$'s, we have
\[
\lim_{t\rightarrow +\infty}\frac{P(T_{\text{EN-Burden}}>t)}{P(V_0>t)} = 1
\]
under the null hypothesis $H_0$, where $V_0$ denotes a standard Cauchy variable.
\end{theorem}
Theorem~\ref{Thm:typeI:burden} implies that the type I error of the ensemble Burden test based on the Cauchy approximation~\eqref{eq:pval:approx} should be well controlled as the significance level goes to 0.

For any true $\boldsymbol{\beta}$ in the alternative $H_1$ in~\eqref{eq:test:problem}, the two-sided oracle test based on the absolute value of $T_\beta = \mathbf{w}_\beta^T\mathbf{S}$ is Bahadur optimal. The following theorem shows the Bahadur efficiency of the ensemble Burden test $T_{\text{EN-Burden}}$ with respect to the theoretical optimal oracle test $T_\beta$. The proof is given in the supplementary materials.

\begin{theorem}\label{Thm:burden}
  (i) For any $\epsilon >0$ and $\delta>0$, by letting $B$ be sufficiently large, we have
  \[
      \text{Eff}_b(T_{\text{EN-Burden}},T_\beta) > 1 - \epsilon
  \]
  with high probability $1-\delta$. (ii) For $b_\beta = \mathbf{w}_\beta^T \boldsymbol{\Sigma} \mathbf{w}_\beta >0$ and sufficiently small $\epsilon >0$, the above result holds when $B\geq O[ \log(1/\delta) (p-1)^{1/2} \{\lambda_{max}/(\epsilon b_\beta)\}^{(p-1)/2}  ]$, where $\lambda_{max}$ is the largest eigenvalue of $\boldsymbol{\Sigma}$.
\end{theorem}

Theorem~\ref{Thm:burden} indicates that the Bahadur efficiency $\text{Eff}_b(T_{\text{EN-Burden}},T_\beta)$ could be infinitely close to 1. As $T_\beta$ is Bahadur optimal,
this means that our proposed ensemble Burden test is near Bahadur optimal with a high probability. As indicated by Theorem~\ref{Thm:burden} (ii), the number of base tests $B$ required to achieve the Bahadur efficiency would increase as the dimension $p$ increases. However, this does not imply that our proposed test would be computationally burdensome for large-scale multiple testing. The computational efficiency of the ensemble tests will be discussed in Section~\ref{Secsub:compu}.
Note that Theorem~\ref{Thm:burden} essentially poses no assumption on the covariance matrix $\boldsymbol{\Sigma}$. Hence, the ensemble Burden test is at least as efficient as the original Burden test in the Bahadur sense for any $\boldsymbol{\Sigma}\geq 0$.

\begin{remark}\label{Rmk1}
  We provide two technical comments about Theorem~\ref{Thm:burden} in this remark.
  (i) $b_\beta = 0$ implies the mean vector of $\mathbf{S}$ is $\mathbf{0}$, i.e., the null is true. (ii)~\cite{littell1973asymptotic} investigated the Bahadur efficiency of the Fisher's combination method for combining independent tests in a meta-analysis setting. The optimal Bahadur efficiency there is to compare the Fisher's method with a class of p-value combination methods, but the resulting test may not be Bahadur efficient for the corresponding alternative hypotheses. In contrast, the Bahadur efficiency in our Theorem~\ref{Thm:burden} shows the ensemble test is Bahadur efficient for  a specific class of alternative hypotheses. Further, unlike~\cite{littell1973asymptotic},  our Theorem~\ref{Thm:burden} concerns correlated tests instead of independent tests, as random weights are applied to the same individual test statistics.
\end{remark}


\subsection{Determine the number of base tests $B$}\label{Secsub:B}
 We note that the discussions in Sections~\ref{Secsub:B}--\ref{Secsub:compu} apply to the general ensemble testing framework introduced in Section~\ref{Subsec:framework}, although the ensemble Burden test is used as a concrete example for the discussions.

\begin{figure}[!h]
  \centering
  \includegraphics[scale=0.65]{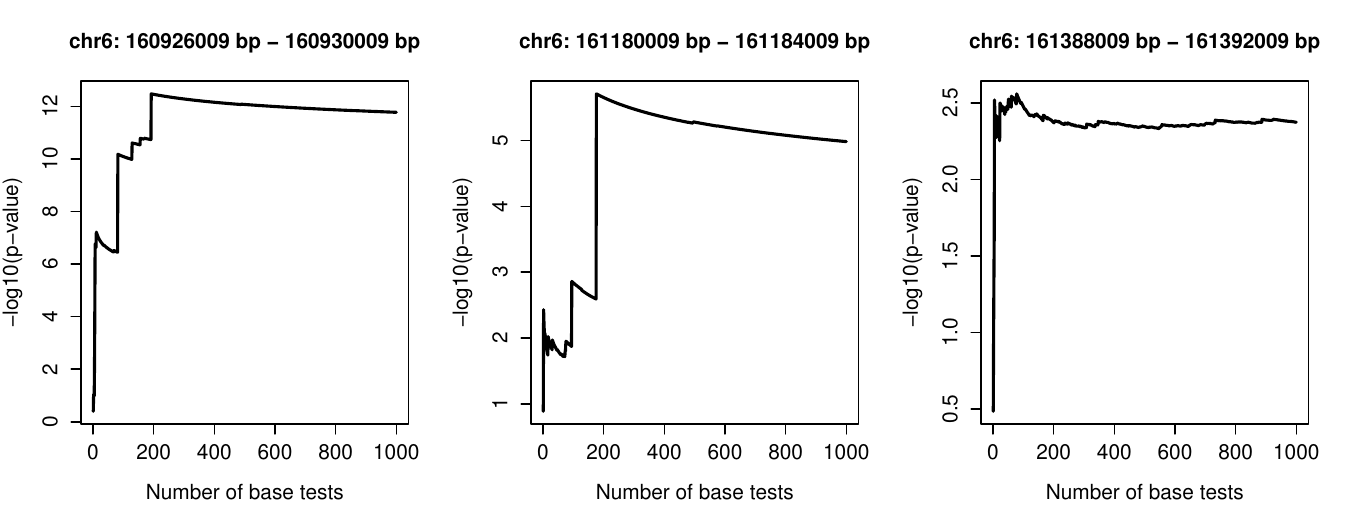}
  \caption{Examples of the p-value path of the ensemble Burden test. The examples are from the analysis of the ARIC whole-genome sequencing data described in Section~\ref{Sec:realdata}. The analysis is for lipoprotein(a) among African Americans. The title of each panel shows the location of the genomic region of SNVs. The numbers of SNVs (i.e., the dimension $p$) in the three examples are 47, 52 and 63, respectively.}
  \label{Eg:pvalpathplot}
\end{figure}

We introduce a general way to determine the number of base tests $B$.
In random forests, one can draw a plot showing how the prediction error (e.g., out-of-bag error) changes as the number of trees $B$ increases~\citep[see, e.g., ][p.p. 591-592]{hastie2009elements}. When the prediction error stabilizes, one can stop growing new trees and output the final random forest. For our ensemble testing method, we can use a similar way to determine the number of base tests $B$. Here, we draw a path of the ensembled p-values $p_{\text{en}}$'s at the $-\log_{10}$ scale as the number of base tests increases and choose $B$ at which $p_{\text{en}}$ becomes stable. Figure~\ref{Eg:pvalpathplot} shows examples of such a plot from the rare variant association analysis of the ARIC WGS data described in Section~\ref{Sec:realdata} for the ensemble Burden test.
Note that the ensemble test statistic $T (\mathcal {D})$ in~\eqref{eq:ETstatistic} is the mean of the transformed p-values $h_{\text{acat}} \{ p(\mathcal {D}, \Theta_i)\}$, and conditional on $\mathcal {D}$, $h_{\text{acat}} \{ p(\mathcal {D}, \Theta_i)\}$'s are i.i.d. random variables. Assume that $h_{\text{acat}} \{ p(\mathcal {D}, \Theta_i)\}$'s are bounded conditional on $\mathcal {D}$. This assumption would hold in most situations, which implies based on the observed data in finite samples, we cannot obtain a p-value of 0 (i.e., completely support the alternative) or 1 (i.e., no evidence at all to support the alternative) for any $\Theta_i$.
Then, by the law of large numbers, $T(\mathcal {D})$ will converge almost surely conditional on $\mathcal {D}$ when $B$ goes to infinity. This guarantees the convergence of $p_{\text{en}}$.

\subsection{Boosting power via variability reduction}\label{Secsub:explanation}


We provide an intuitive explanation about why the ensemble approach could be helpful for developing powerful tests, which shares some similarities with that for the random forest.
Trees can capture complicated nonlinear structures and have low bias if they grow sufficiently deep~\citep{hastie2009elements}. But they are notoriously unstable and have a high variance~\citep{breiman1996bagging}. Through the aggregation of trees (i.e., base learners), random forest reduces the variance and hence improves the performance~\citep[e.g.,][]{buhlmann2002analyzing}. Here, the superior performance of ensemble tests can be explained in a similar fashion via certain kind of variability reduction.

As mentioned in Section~\ref{Sec:1}, since the alternative hypothesis is composite and consists of different specific alternatives (i.e., values of $\boldsymbol{\beta}$), the UMP test does not exist and any test could perform poorly under a wide range of alternatives~\citep[][Chapter 14.6]{lehmann2006testing}. Hence, it is not reasonable to require a test to perform well against all alternatives, and roughly speaking, a strong test would mean the test has a high power against as many alternatives in a class of alternatives as possible.

Consider the base tests in our proposed ensemble Burden test. If the true signal direction $\mathbf{w}_\beta$ happens to be the same as or close to the randomly  chosen $\mathbf{w}_i$ in the base test $T_i$, then $T_i$ would have the best power. However, the base test is vulnerable to the deviation of its chosen direction from  the  true signal direction, and  may perform poorly even if $\mathbf{w}_\beta$ deviates slightly from $\mathbf{w}_i$ in the presence of negative correlation among $\mathbf{X}$ (see Figure~\ref{Fig:Burden_limitation}(d) and suppose $\mathbf{w}_i = \mathbf{1}_p$ for an example).  In other words, the power of the base test is unstable and has a high  variability across different alternatives. In practice, we do not know $\mathbf{w}_\beta$ or the underlying true alternative. Therefore, individual base tests are weak.

 To improve over the base test, we would like to obtain a strong test in such a way that dampens the impact of those base tests that have low powers (e.g., $\mathbf{w}_\beta$'s far away from $\mathbf{w}_i$), but features
the base tests that are powerful (e.g., $\mathbf{w}_\beta$'s close to $\mathbf{w}_i$).  In this way, we not only reduce the power variability of base tests (i.e., making base tests stable), but also achieve strong or high power across different specific alternatives. Some graphical illustrations of these points using the comparison of the base and ensemble Burden tests are provided in Section 4 in the Supplementary Materials. Through reducing the variability of power across different alternatives (i.e., $\mathbf{w}_\beta$'s), the ensemble Burden test provides a substantial increase in the average power, and in the meantime becomes more stable. Hence, the ensemble Burden test is a strong test.

\begin{figure}[!h]
  \centering
  \includegraphics[scale=0.55]{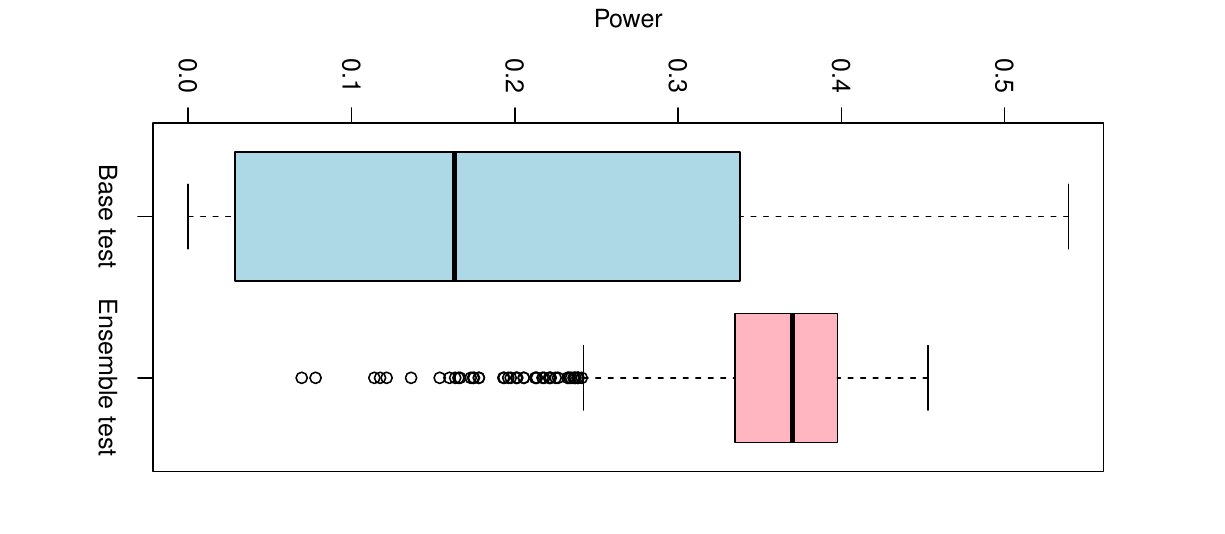}
  \caption{Powers of the base and ensemble Burden tests at 1000 different parameter values under the alternative, summarized by boxplots. The detail of this example is described in Section 4 in the Supplementary Materials.}
  \label{Eg:boxplot:eg}
\end{figure}

Note when calling a test is weak or strong, here we consider the  overall performance (or power) of a test under the composite alternative (which consists of many different specific alternatives) instead of the power at a fixed alternative. In this sense, all the base tests are weak. Figure~\ref{Eg:boxplot:eg} shows an example that demonstrates the better overall performance (or power) of the ensemble Burden test compared to a base Burden test. In order to be powerful to detect a true signal, our ensemble test does need some random base tests to have a high power against the alternative of unknown true signal, as indicated by Theorem 2.

\subsection{Reasons to use ACAT for combining base tests}\label{Secsub:ACAT}

  Our proposed ensemble framework uses ACAT~\citep{liu2019acat,liu2020cauchy} to aggregate the  p-values of many base tests, which are correlated as they are constructed using the same data by incorporating certain  random procedure (e.g., random weights). ACAT provides a computationally highly efficient and powerful approach to combine correlated p-values. First, the final combined p-value of ACAT can be simply approximated by~\eqref{eq:pval:approx} based on a Cauchy distribution. This approximation is particularly accurate when $p_{\text{en}}$ is very small~\citep{liu2020cauchy}, which is well-suited for the computationally challenging situations of large-scale multiple testing (see also Section~\ref{Subsec:challenges:large:testing}). Second, since a wide range of base tests are explored and many base tests could only have litter powers against the alternative of the unknown true signal, we would like to aggregate the base tests in a way such that those underpowered base tests only play a small role in the aggregation. As ACAT mainly uses several smallest p-values of the base tests to represent the overall significance~\citep{liu2019acat}, the underpowered base tests would only have limited impact on the final ensemble test.

\subsection{Computational efficiency of the ensemble testing framework}\label{Secsub:compu}

As mentioned in Section~\ref{Subsec:challenges:large:testing}, we aim to develop tests that are computationally scalable in the large-scale multiple testing setting.  Here, we discuss the computational efficiency of our ensemble testing framework from four aspects, i.e., the method for p-value combination, computational implementation, choices of base tests, and the number of base tests $B$.

Firstly, as discussed in Section~\ref{Secsub:ACAT}, the ACAT method used in our framework is a computationally highly efficient way to combine the p-values of base tests. Secondly, since the random components $\Theta_i$'s in our framework are i.i.d., the p-values of the base tests can be computed simultaneously.
Hence, it is straightforward to implement the ensemble tests via parallel computing to reduce the computation time for applications that are computationally demanding. Thirdly, we typically select existing popular tests (e.g., the Burden test) as the base tests in our framework, as they often have a low computational requirement.

Lastly, we discuss the computational burden regarding the number of base tests $B$ required by the ensemble test.
First of all, in applications that involve large-scale multiple testing, while the total number of tests could be huge, the dimension $p$ in the individual hypotheses is typically not large. For instance, although millions of variant sets need to be tested in WGS studies,  each variant set often only contains a moderate number of variants, e.g., $p$ is often in tens or hundreds. In the situations of a non-large $p$, our empirical studies (e.g., the examples in Figure~\ref{Eg:pvalpathplot} and the ARIC data analysis in Figure~\ref{Fig:SigWinPath}) demonstrate that in practice the ensembled p-value $p_{\text{en}}$ becomes stable when $B$ is moderate.

 Second, as described in Section~\ref{Subsec:challenges:large:testing}, the main computational burden in the large-scale multiple testing setting is not from a high dimension $p$, but from the need to apply a test a massive number of times. If the ensemble test requires a large $B$ every time, it would be computationally burdensome. However, by monitoring the path of the ensemble test p-value $p_{\text{en}}$ for each variant set, one does not need to wait until $p_{\text{en}}$ is stable every time and one can stop increasing $B$ early in most times. For instance, suppose that the genome-wide significance threshold is $\alpha = 10^{-8}$ and when $B = 300$, the $p_{\text{en}}$ of a variant set fluctuates around $10^{-2}$. As this variant set has no hope to pass the significance threshold, there is no need to continue adding more base tests and one can stop increasing $B$. Since a vast majority of variant sets across the genome are from the null, one can use a moderate $B$ for a vast majority of variant sets.  In addition, another situation where one can stop earlier is when a variant set is super significant, e.g., $p_{\text{en}} = 10^{-12}$ when $B = 300$ and there is no need to increase $B$ further. Figure~\ref{Eg:pvalpathplot} provides examples of these two situations.
Lastly, in the presence of a very high dimension $p$, the ensemble tests proposed in Sections~\ref{Sec:Burden}--\ref{Sec:chisq} would require a larger $B$. We discuss a possible way to reduce $B$ in Section~\ref{Sec:diss}.

\section{Extensions of the ensemble Burden test with auxiliary information about relative effect sizes}\label{Sec:aux:info}

For variant set tests in WGS studies, based on genetic knowledge, we can leverage some auxiliary information to obtain prior knowdedge about the effect sizes of SNVs, i.e., the relative magnitudes of $\beta_j$'s, to further empower the analysis. For example, the genetic effect sizes of SNVs are expected to be related to their minor allele frequencies (MAFs)~\citep{wu2011rare} and their functional annotations~\citep{li2020dynamic}. Let $a_j$ denote the relative magnitude of $\beta_j$ based on the prior knowledge, where $j=1,2,\cdots,p$. The Burden test incorporates the prior information by weighting the marginal score statistics $S_j$'s by the relative magnitudes and summing up the weighted $S_j$'s, i.e., $T_{\text{Burden}} = \mathbf{a}^T\mathbf{S}$, where $\mathbf{a} = (a_1,a_2,\cdots,a_p)^T$.
For instance, \cite{wu2011rare} proposed to set  $a_j= Beta(MAF_j,1,25)$  to upweight SNVs with smaller MAFs, where $Beta(MAF_j,1,25)$ denotes the beta density function with parameters 1 and 25 evaluated at the MAF of the $j$-th SNV.

Suppose that $a_j > a_k$, where $a_j$ and $a_k$ are the $j$-th and $k$-th components of $\mathbf{a}$, respectively. Typically, this means that we know $|\beta_j|$ are likely to be larger than $|\beta_k|$, but does not mean $|\beta_j|$ must be greater than $|\beta_k|$. For instance, SNVs with smaller MAFs generally have stronger effects than those with larger MAFs, but this negative relationship between MAF and effect size is not always true for any two SNVs. However, the Burden test assumes a deterministic relationship between $a_j$ and $|\beta_j|$, which is not the case in reality and therefore could lead to power loss. A more appropriate assumption to characterize the scenario with the auxiliary information would be that the variables have the same effect sign and their expected effect sizes are proportional to $a_j$'s.

It would be challenging to develop powerful tests in this scenario via the traditional approaches. In contrast,
constructing an ensemble test to characterize the scenario with the auxiliary information is a simple and straightforward extension of what was presented in Section~\ref{Secsub:ENBurden}. Recall that the random weights are generated using multivariate Gaussian distribution with covariance matrix $\mathbf{I}$. To incorporate the auxiliary information, we only need to replace the identity matrix $\mathbf{I}$ by $\mathbf{A}^2$, where $\mathbf{A}$ is a diagonal matrix with its diagonals being $a_j$'s. Specifically, the random weights are generated by
\begin{equation}\label{Eq:random:weight:generate}
  \mathbf{w}_i  = \left(\frac{|\xi_{i1}|}{||\boldsymbol{\xi}_i||},\frac{|\xi_{i2}|}{||\boldsymbol{\xi}_i||},\cdots,\frac{|\xi_{ip}|}{||\boldsymbol{\xi}_i||}\right)^T, \quad \boldsymbol{\xi}_i \sim N_p(0,\mathbf{A}^2).
\end{equation}
The rest of ensemble testing procedure proceeds as usual. We can see that the ratio of the weight expectations for any two variables (i.e., $E[w_{ij}]/E[w_{ik}]$) is the same as the ratio of their effect sizes (i.e., $a_j/a_k$). In this way, the variable with an expected larger effect size is more likely to receive a higher weight and therefore plays a more important role in the ensemble test.

\section{The Ensemble SKAT and MORST Tests}\label{Sec:SKAT}

\subsection{Ensemble SKAT}\label{Secsub:ENSKAT}

In WGS variant set analysis, another widely used test is the Sequence Kernel Association Test (SKAT)~\citep{wu2011rare}, which does not assume $\beta_j$'s have the same sign. The basic model for SKAT is the same as model (\ref{eq:model}) used for the Burden test introduced in Section~\ref{Secsub:basic:Burden}, except that the alternative hypothesis assumes the signals are weak and may be in different directions. SKAT is more powerful for detecting variant sets with different effect signs, while the Burden test is more advantageous for variant sets with the same effect sign~\citep{lee2014rare}.
	
 Specifically, SKAT imposes a working assumption that $\beta_j$'s are independent and follow an arbitrary distribution with mean 0 and variance $\tau$. Then, testing for $H_0: \boldsymbol{\beta}=0$ vs $H_1:\boldsymbol{\beta}\neq 0$ is equivalent to test the variance component $H_0: \tau=0$ vs $H_1:\tau>0$.  SKAT uses a score statistic to test for $H_0: \tau=0$, which is
 a quadratic combination of $\mathbf{S}$, i.e., $T_{\text{SKAT}} = \mathbf{S}^T\mathbf{W}^T\mathbf{W}\mathbf{S}$, where $\mathbf{W}$ is a diagonal matrix of weights. In comparison, the Burden test statistic is a (weighted) linear combination of $\mathbf{S}$. Since $T_{\text{SKAT}}$ follows a non-standard distribution (i.e., a mixture of chi-squared distributions), it is mathematically difficult to derive the finite-sample power of SKAT in a closed form. Therefore, similar to the limitation of the Burden test, while equal weight seems to be a natural choice for SKAT if there is no prior information about the effect sizes $|\beta_j|$'s, it may not give the optimal power because of the covariance $\boldsymbol{\Sigma}$ of the marginal score statistics. A discussion on the limitation of SKAT is also provided in Section 8 in the Supplementary Materials. Since the best choice of weights $\mathbf{W}$ is unknown under a specific $\boldsymbol{\Sigma}$ in finite samples, the discussions on the power variability reduction described in Section~\ref{Secsub:explanation} suggest that we can explore a variety of  different $\mathbf{W}$'s for SKAT using random weights to obtain a powerful and computationally scalable ensemble test.

 To be specific, we use exactly the same way of constructing the ensemble Burden test to develop an ensemble SKAT test to boost power for SKAT. Let $\mathbf{W}_i$ be a diagonal matrix with the diagonal being a vector of randomly simulated weights $\mathbf{w}_i$, where $i=1,2,\cdots,B$. Since SKAT does not assume the effect signs are the same, we should generate the $\mathbf{w}_i$'s independently from a uniform distribution on $\mathcal{S}_p = \{ (w_1,w_2,\cdots,w_p): \sum_{j=1}^p w_j^2 = 1 \}$. Note that the weights in SKAT are squared. It is therefore equivalent to
generate $\mathbf{w}_i$'s independently from a uniform distribution on $\mathcal{S}_p^+$. The SKAT statistic for the $i$-th base test is given by
\begin{equation*}
   T_{i,\text{SKAT}} = \mathbf{S}^T\mathbf{W}_i^T\mathbf{W}_i\mathbf{S}.
\end{equation*}
 Let $\lambda_{i1}\geq \lambda_{i2}\geq, \cdots, \geq\lambda_{ip}\geq 0$ be the eigenvalues of $\mathbf{W}_i\boldsymbol{\Sigma} \mathbf{W}_i$. Under the null, $T_{i,\text{SKAT}} \sim \sum_{j=1}^p \lambda_{ij} \chi_j^2(1)$, where $\chi_j^2(1)$'s denote i.i.d. chi-squared random variables with 1 degree of freedom. The ensemble SKAT statistic is
\begin{equation*}
   T_{\text{EN-SKAT}} = \frac{1}{B} \sum_{i=1}^B \tan\{ [F_i(T_{i,\text{SKAT}})-1/2]\pi \}.
\end{equation*}
where $F_i(\cdot)$ be the c.d.f. of $\sum_{j=1}^p \lambda_{ij} \chi^2_j(1)$.
The last step is to approximate the p-value of $T_{\text{EN-SKAT}}$ by the Cauchy-based-approximation~\eqref{eq:pval:approx}.

For the ensemble SKAT, we can establish theoretical results that are parallel to Theorems~\ref{Thm:typeI:burden} and~\ref{Thm:burden} for the ensemble Burden test. The following theorem indicates the type I error of the ensemble SKAT would be controlled when the significance level goes to 0.
\begin{theorem}\label{Thm:typeI:SKAT}
For a fixed $B\geq 1$ and a fixed realization of $\mathbf{w}_i$'s, suppose that $\lambda_{i1}>\lambda_{i2}$ for any $1\leq i \leq B$. Then, under the null hypothesis $H_0$, we have
\[
\lim_{t\rightarrow +\infty}\frac{P(T_{\text{EN-SKAT}}>t)}{P(V_0>t)} = 1,
\]
where $V_0$ denotes a standard Cauchy variable.
\end{theorem}
The condition that $\lambda_{i1}>\lambda_{i2}$ for any $1\leq i \leq B$ is a technical condition. Since $\mathbf{w}_i$'s are randomly generated from $\mathcal{S}_p$, this condition would hold with probability 1. We note that the theoretical results in~\cite{liu2020cauchy} only hold for normally distributed test statistics. As the base SKAT statistics follow mixture of chi-squared distributions, substantial new theoretical developments are needed to establish Theorem~\ref{Thm:typeI:SKAT}. In particular, the main technical developments are about dealing with this class of non-standard distributions (i.e., mixtures of chi-squared distributions), including calculating their tail probabilities and handling the dependency of base test statistics following these distributions (i.e., lemmas 5, 7 and 10 in the supplementary materials). The proof is given in the supplementary materials.

Let $T_{\beta,\text{SKAT}}$ denote the oracle test that uses the true signal direction $\mathbf{w}_\beta = \boldsymbol{\beta}/||\boldsymbol{\beta}||$ to weight the marginal score test statistics. Parallel to Theorem~\ref{Thm:burden}, the following theorem indicates the Bahadur efficiency of the ensemble SKAT could be sufficiently close to the oracle test $T_{\beta,\text{SKAT}}$.
\begin{theorem}\label{Thm:SKAT}
  For any $\epsilon >0$ and $\delta>0$, by letting $B$ be sufficiently large, we have
  \[
      \text{Eff}_b(T_{\text{EN-SKAT}},T_{\beta,\text{SKAT}}) > 1 - \epsilon
  \]
  with high probability $1-\delta$.
\end{theorem}

\begin{remark}\label{Rmk2}
 When $\boldsymbol{\Sigma}$ is a diagonal matrix, similar to Theorem~\ref{Thm:burden} (ii), we can also drive a lower bound of $B$ for the ensemble SKAT, which is provided at the end of the proof of Theorem~\ref{Thm:SKAT} in the Supplementary Materials.
\end{remark}

\subsection{Ensemble MORST}

SKAT is a score test and therefore is locally most powerful~\citep{wu2011rare}. This implies that SKAT is only advantageous under local alternatives with weak signals and could lose considerable power in the presence of moderate or strong signals.
To overcome the limitations of SKAT, \cite{liu2020minimax} proposed a Minimax Optimal Ridge-Type Set Test (MORST) that has robust power across different signal strengthes. The MORST test statistic is $T_{\text{MORST}} = \mathbf{S}^T\mathbf{W}^T (\mathbf{I} + \theta \mathbf{W}\boldsymbol{\Sigma}\mathbf{W}^T )^{-1}  \mathbf{W}\mathbf{S}$, where $\mathbf{W}$ is the same as that in $T_{\text{SKAT}}$, and the parameter $\theta$, which depends only on the eigenvalues of $\boldsymbol{\Sigma}$ and the significance level, is obtained based on a minimax criterion to make the power robust to signal strength.

We construct an ensemble MORST by  incorporating random weights into the test statistic.   Following the notations in Section~\ref{Secsub:ENSKAT}, the MORST statistic for the $i$-th base test is
\begin{equation*}
  T_{i,\text{MORST}} =  \mathbf{S}^T\mathbf{W}_i^T (\mathbf{I} + \theta_i \mathbf{W}_i\boldsymbol{\Sigma}\mathbf{W}_i^T )^{-1}  \mathbf{W}_i\mathbf{S}.
\end{equation*}
Note that $T_{i,\text{MORST}}$ also follows a mixture of chi-squared distributions under the null.
Let $\lambda_{i1}\geq \lambda_{i2}\geq, \cdots, \geq\lambda_{ip}\geq 0$ be the eigenvalues of $(\boldsymbol{\Sigma}^{1/2})^T\mathbf{W}_i^T (\mathbf{I} + \theta_i \mathbf{W}_i\boldsymbol{\Sigma}\mathbf{W}_i^T )^{-1}\mathbf{W}_i\boldsymbol{\Sigma}^{1/2}$ and $F_i(\cdot)$ be the c.d.f. of $\sum_{j=1}^p \lambda_{ij} \chi^2_j(1)$. Then, the ensemble MORST statistic is
\begin{equation*}
   T_{\text{EN-MORST}} = \frac{1}{B} \sum_{i=1}^B \tan\{ [F_i(T_{i,\text{MORST}})-1/2]\pi \}.
\end{equation*}

Since the distribution of MORST statistic under the null or alternative is similar to that of the SKAT statistic as a mixture of chi-square distribution with a different set of eigenvalues,  the theoretical results for the ensemble MORST directly follow from those for the ensemble SKAT. Let $T_{\beta,\text{MORST}}$ denote the oracle test that uses the true signal direction $\mathbf{w}_\beta = \boldsymbol{\beta}/||\boldsymbol{\beta}||$ as the weight vector in $T_{\text{MORST}}$. We have the following corollary about the the ensemble MORST test, which shows its type I error control and Bahadur efficiency relative to the oracle test $T_{\beta,\text{MORST}}$.

\begin{corollary}
  The results in Theorems~\ref{Thm:typeI:SKAT} and~\ref{Thm:SKAT} hold with the ensemble SKAT being replaced by the ensemble MORST.
\end{corollary}

%

\subsection{Extensions with auxiliary information about relative effect sizes}\label{Subsec:incor_beta_SKAT_MORST}
SKAT and MORST use the same way as the Burden test does to incorporate the auxiliary information about relative effect sizes, i.e., by setting the weights to be $a_j$'s directly as described in Section~\ref{Sec:aux:info}. Hence, for ensemble SKAT or MORST, we can also generate the random weights according to~\eqref{Eq:random:weight:generate} to better leverage the auxiliary information and improve power.

%

\section{The Ensemble Subset Chi-squared Test}\label{Sec:chisq}

In this section, we apply the proposed ensemble framework to study another testing problem, i.e., testing against sparse alternatives. The notations in this section are self-contained.

\subsection{The sparse signal setting}\label{Subsec:sparse:testing}

In many contemporary applications, it becomes increasingly easy and cheap to collect a large number of variables, among which, however, only a small proportion of them are expected to be relevant. Such a situation is often referred to as sparse signals or signal sparsity. How to incorporate the sparsity information to improve testing power has received substantial interest in the literature~\citep[see, e.g.,][]{ingster1996some,donoho2015special}. Here, to fix the idea, we consider the basic multivariate normal model
\begin{equation}\label{eq:normal:model}
    \mathbf{Z} \sim N_p(\boldsymbol{\mu},\boldsymbol{\Omega}),
\end{equation}
where $\mathbf{Z}=(Z_1,Z_2,\cdots,Z_p)^T$ is a $p$-dimensional Gaussian random vector, $\boldsymbol{\mu}=(\mu_1,\mu_2,\cdots,\mu_p)^T$ is the mean vector, and $\boldsymbol{\Omega}$ is a known correlation matrix.
The question of interest is to test
\begin{equation}\label{eq:test:sparse:problem}
  H_0: \boldsymbol{\mu}= 0  \quad\quad \text{against}  \quad \quad H_1: \boldsymbol{\mu} \neq 0 \;\text{and there are only} \; m \ll p \; \text{nonzero}\; \mu_j\text{'s},
\end{equation}
where $m$ denotes the number of signals and is much smaller than the dimension $p$. When $\boldsymbol{\Omega}$ is an identity matrix or some weak-dependence matrix, the asymptotic theory about the detection boundary has been well studied~\citep{ingster1996some,donoho2004higher,hall2010innovated}, which characterizes the relationship of the dimension $p$, the number of signals $m$, and the signal strengths $\mu_j$'s for situations where detection is possible and impossible when $p$ goes to infinity.  Extensions of the results for model~\eqref{eq:normal:model} to other settings have also been studied, e.g., the linear regression setting~\citep{arias2011global}. In particular, extensions to the variant set test setting in Section~\ref{Sec:Burden} based on the score statistics $\mathbf{S}$  in  genetic   association studies can be found in~\cite{barnett2017generalized} and  \cite{sun2020genetic}.

\subsection{Existing tests}
Among tests that are designed to be particularly powerful under sparse alternatives, the Higher Criticism (HC) ~\citep{donoho2004higher} and Berk-Jones (BJ) tests~\citep{berk1979goodness} have received considerable attention in the literature.
When $\boldsymbol{\Omega}$ in~\eqref{eq:normal:model} is an identity matrix,  \cite{donoho2004higher} showed that both the HC and BJ tests achieve the asymptotic detection boundary, which also means the two tests are asymptotically minimax optimal in the sparse regime. However, in nonasymptotic situations where the dimension $p$ is not very large, as the case in varaint set tests in WGS studies, the powers of the two tests are often substantially different~\citep[see, e.g.,][]{sun2020genetic}. Specifically, the HC test typically has a higher power than the BJ test in the presence of extremely sparse signals, and a lower power in the presence of moderately sparse signals.
To efficiently handle the cases with correlated $Z_j$'s, several extensions of the HC and BJ tests have been proposed, such as the innovated higher criticism~\citep{hall2010innovated}, generalized higher criticism~\citep{barnett2017generalized}, and generalized Berk-Jones tests~\citep{sun2020genetic}. When $p$ is not super large, the powers of these extensions are also considerably distinct  under different levels of signal sparsity.

Overall, for a non-large $p$, in both independent and correlated cases, it is not clear whether the HC, BJ and their extensions would have the best power in practice. In addition, calculations of p-values of these tests are computationally intensive even if $p$ is moderate~\citep{barnett2014analytical,barnett2017generalized, sun2020genetic}. Since the true signal variants in a set are unknown in practice, following the discussions on the power variability reduction in Section~\ref{Secsub:explanation}, we explore the ensemble strategy by randomly selecting a subset of individual variants, and  develop a computationally scalable ensemble test that have more robust power than these existing tests across various levels of sparsity.

\subsection{Ensemble subset chi-squared test}

We employ the ensemble framework described in Section~\ref{Subsec:framework} to construct tests for sparse alternatives~\eqref{eq:test:sparse:problem}. For now, assume that the number of signals $m$ is known. First, we randomly draw $m$ indices from $\{1,2,\cdots, p\}$ without replacement and repeat this process independently to obtain a collection of index sets, where the $i$-th index set is denoted by $\mathcal{J}_i$ and $i=1,2,\cdots,B$. In other words, $\mathcal{J}_i$'s are random and independent subsets of $\{1,2,\cdots, p\}$ with cardinality $m$.
Define the base test statistic as
\begin{equation*}
   T_i = \mathbf{Z}_{\mathcal{J}_i}^T \boldsymbol{\Omega}_{\mathcal{J}_i}^{-1} \mathbf{Z}_{\mathcal{J}_i},
\end{equation*}
where $\mathbf{Z}_{\mathcal{J}_i}$ is a subvector of $\mathbf{Z}$ and $\boldsymbol{\Omega}_{\mathcal{J}_i}$ is a submatrix of $\boldsymbol{\Omega}$ corresponding to the indices in $\mathcal{J}_i$. Under the null, the test statistic $T_i$ follows $\chi^2(m)$, which denotes a chi-squared distribution with $m$ degrees of freedom. Let $q_m(\cdot)$ be the c.d.f. of $\chi^2(m)$. Then, we combine the p-values of the base tests $T_i$'s via ACAT and obtain the ensemble subset chi-squared test statistic
\begin{equation*}
  T_{\text{EN-Subset-Chisq}} = \frac{1}{B}\sum_{i=1}^B \tan\{ [q_m(T_i)-1/2]\pi \}.
\end{equation*}
The final step is to approximate the p-value of $T_{\text{EN-Subset-Chisq}}$ by~\eqref{eq:pval:approx}. It can be seen that the random component in the ensemble subset chi-squared test, i.e., the random selection of variables, characterizes the scenario of sparse signals.

The rationale of the ensemble subset chi-squared test can be similarly explained to that of the ensemble Burden test described in Section~\ref{Secsub:explanation}.
Let $\mathcal{J}_\mu = \{j: \mu_j\neq 0\}$ denote the set of signals. If $\mathcal{J}_i = \mathcal{J}_\mu$, the chi-squared test based on $T_i$ would be very powerful as it includes all the signals without adding any noises. But if $\mathcal{J}_i$ contains no or few signals, $T_i$ would have no or little power. Therefore, the power of $T_i$ has a high variability across the sparse alternatives in~\eqref{eq:test:sparse:problem}. The aggregation of the base tests reduces the variability and make the power of the final ensemble test robust to the various alternatives in $H_1$ of sparse signals.

We investigate the theoretical properties of the ensemble subset chi-squared test in terms of Bahadur efficiency that was briefly introduced in Section~\ref{Secsub:Bahadur}.
As with the ensemble Burden test, we first show the null distribution of $T_{\text{EN-Subset-Chisq}}$ is a standard Cauchy in the tail.
\begin{theorem}\label{Thm:typeI:ENchisq}
For any fixed $B\geq 1$ and any fixed realizations of $\mathcal{J}_i$'s, we have
\[
\lim_{t\rightarrow +\infty}\frac{P(T_{\text{EN-Subset-Chisq}}>t)}{P(V_0>t)} = 1
\]
 under the null hypothesis $H_0$, where $V_0$ denote a standard Cauchy variable.
\end{theorem}
 Define $T_{\mu} = \mathbf{Z}_{\mathcal{J}_\mu}^T \boldsymbol{\Omega}_{\mathcal{J}_\mu}^{-1} \mathbf{Z}_{\mathcal{J}_\mu}$, where $\mathcal{J}_\mu= \{j: \mu_j\neq 0\}$ indicates the true set of signals. We refer to $T_\mu$ as the oracle test since it only uses $Z_j$'s that have non-zero means to construct the chi-squared test. The following theorem provides the Bahadur efficiency of $T_{\text{EN-Subset-Chisq}}$ with respect to the oracle test $T_\mu$.
\begin{theorem}\label{Thm:ENchisq}
  (i) For any $\epsilon >0$ and $\delta>0$ and a fixed but sufficiently large $B$, we have
  \[
      \text{Eff}_b(T_{\text{EN-Subset-Chisq}},T_\mu) > 1 - \epsilon
  \]
  with high probability $1-\delta$.  (ii) The result in (i) holds when $B \geq O(p^m \log(1/\delta))$.
\end{theorem}

Let $s$ be the number of variables in the base subset chi-squared test $T_i$. We have selected $s = m$ if the number of signals $m$ is known. Next, we discuss the case where $m$ is unknown. There could be many possible ways to deal with an unknown $m$, such as estimating $m$ or further selecting $s$ randomly in the ensemble test. As the focus here is to demonstrate the benefits of ensembling, we provide a simple way based on the theoretical threshold that separates the sparse regime from the dense regime \citep{donoho2004higher}. But we note that the selection of $s$ warrants further study and is a future research direction.

The theoretical analysis in \cite{donoho2004higher} indicates that signals are considered as sparse if $m \leq \sqrt{p}$ or dense if $m>\sqrt{p}$ in the asymptotic sense where $p$ goes to infinity.  As we would like the ensemble subset chi-squared test to be powerful against sparse alternatives, we choose $s$ to be the threshold that separates the sparse and dense regimes, i.e., $s=[\sqrt{p}]$. While if $m<s=\sqrt{p}$, our choice of $s$ is sub-optimal. The power loss is limited because the chi-squared test does not require the number of signals to be the same as  the degree of freedom of the chi-squared test in order to be powerful. In fact, the theory in \cite{donoho2004higher} also indicates that  as long as the number of signals is greater than the square root of the degree of freedom (i.e., dense signals), the chi-squared test would have a high power in an asymptotic sense.   Therefore, we choose $s$ as the upper bound of the number of signals in the sparse regime.

\begin{figure}[!h]
  \centering
  \includegraphics[scale=0.45]{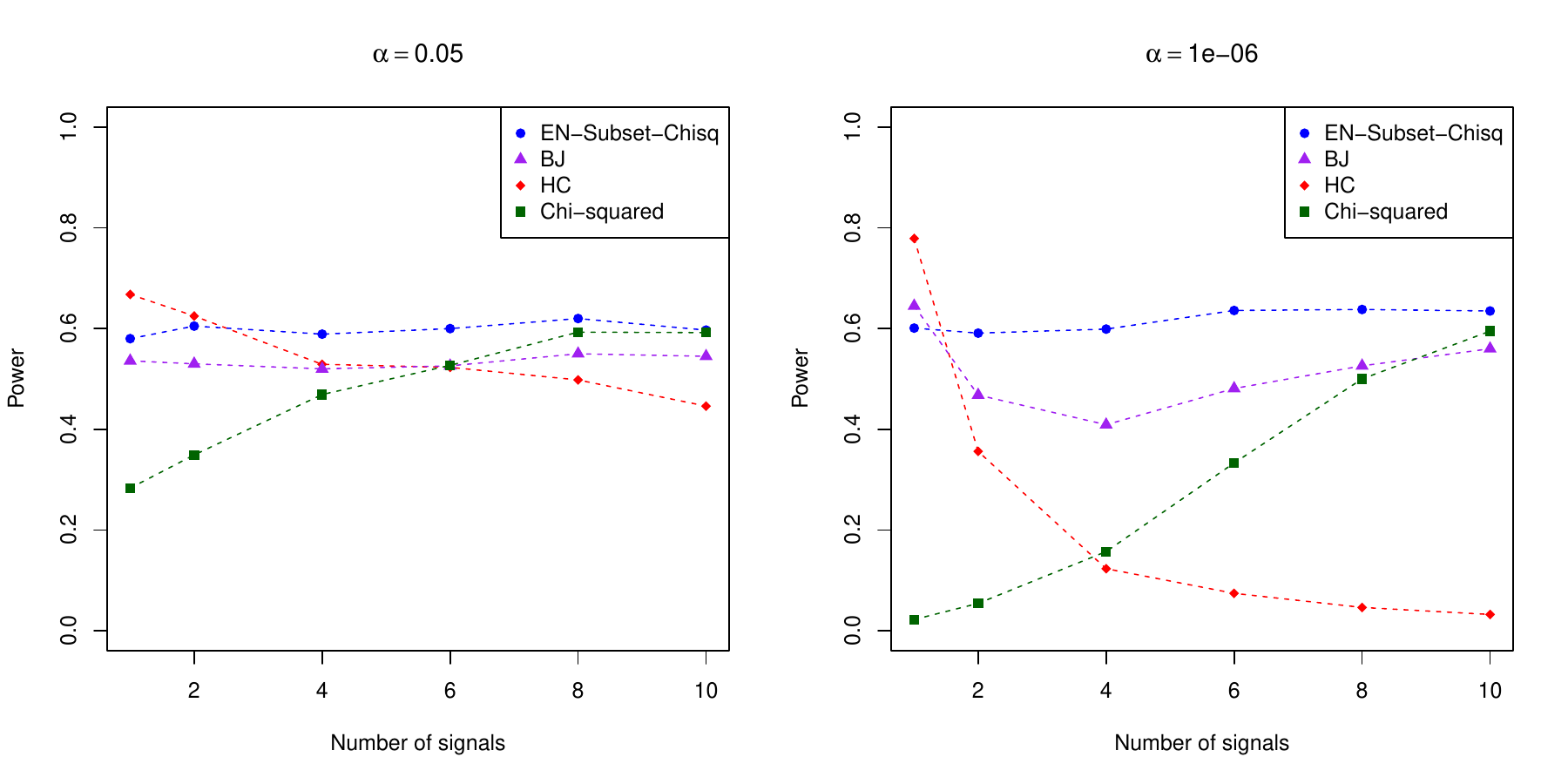}
  \caption{Power comparison of the ensemble subset chi-squared test, the Berk-Jones test, the Higher Criticism test, and the traditional chi-squared test with $p$ degrees of freedom in a multivariate Gaussian model. Specifically, $\mathbf{Z} \sim N_p(\boldsymbol{\mu},\mathbf{I})$ , where $p=100$. The number of signals (i.e., non-zero $\mu_j$'s) is set to be $m = 1,2,4,6,8,10$. The signals have a common signal strength $\mu_0$. In each scenario, $\mu_0$ is chosen such that the power of ensemble subset chi-squared test is around 0.6. In the ensemble subset chi-squared test, the number of variables in the base test is set as $s=10$ and the number of base tests is $B=10^4$. Two significance levels are considered: $\alpha = 0.05, 10^{-6} $. When $\alpha=0.05$, the critical values of the ensemble subset chi-squared test are obtained through simulations to make a fair comparison. The power of each test is calculated based on 1000 replications. }
  \label{PowerExampleENchisq}
\end{figure}

 Empirically, we observe this choice of $s$ performs well when $p$ is not very large. Figure~\ref{PowerExampleENchisq} displays  simulation results under model~\eqref{eq:normal:model} with $p=100$, which compares the power of the ensemble subset chi-squared test with $s=\sqrt{p}$, HC, BJ and the traditional chi-squared test with $p$ degrees of freedom when $m$ varies. Even if $s$ is misspecified (i.e., $s \neq m$), the ensemble subset chi-squared test has a higher power than HC and BJ unless in the extremely sparse case where $m=1$. Similar to the ensemble Burden test, the power gain of the ensemble subset chi-squared test is more significant at a smaller $\alpha$. The traditional chi-squared test has a limited power in the presence of sparse signals but becomes as powerful as the ensemble subset chi-squared test when $m=\sqrt{p}$. More simulation results about the ensemble subset chi-squared test are provided in the Supplementary Materials, which also demonstrate its superior performance.

\section{Simulation studies}\label{Sec:simu}

In this section, we conduct simulations to evaluate the type I errors of the ensemble Burden, SKAT and MORST tests, and compare their powers with the original Burden, SKAT and MORST, respectively. Due to space limitation, the simulation studies and a real-data analysis for the ensemble subset chi-squared test are provided in Sections 2 and 3 in the Supplementary Materials.

We generate continuous responses by
\[
    Y = 0.5Z_1+0.5Z_2+\mathbf{G}^T\boldsymbol{\beta} + \epsilon,
\]
where $Z_1$ is a continuous covariate simulated from $N(0,1)$, $Z_2$ is a binary covariate taking values of -0.5 and 0.5 with equal probability, $\mathbf{G}$ is the genotypes of $p$ SNVs, $\boldsymbol{\beta}$ is a vector of regression coefficients, and $\epsilon$ is an error term following $N(0,1)$. We use the genetic software COSI~\citep{schaffner2005calibrating} to simulate 100 sets of 1 Mb big regions of sequencing data that mimics the correlation structures of sequenced genotypes in the European population. The genotypes $\mathbf{G}$ are a randomly selected continuous subregion with $p=100$ SNVs from one of the 100 big regions. Throughout the simulation, the sample size $n$ is 10000.

In our simulations, we also consider the cases with auxiliary information about the effect sizes of SNVs (i.e., the magnitudes of $\beta_j$'s). As discussed in Section~\ref{Sec:aux:info}, \cite{wu2011rare} proposed to set $a_j = Beta(MAF_j,c_1,c_2)$, which denotes the density of a beta distribution with parameters $c_1$ and $c_2$ evaluated at the MAF of the $j$-th SNV, and then used $a_j$'s as the weights in the Burden test and SKAT. Common choices of parameters include $c_1=1$ and $c_2=25$ (i.e., $Beta(1,25)$), which implies a negative relationship between MAF and effect size, and $c_1=c_2=1$ (i.e., $Beta(1,1)$), which means the MAF and effect size are unrelated. Note that $Beta(1,1)$ corresponds to the case where no prior information about effect sizes is used, while $Beta(1,25)$ corresponds to the situation where rarer variants are expected to have larger effect sizes as discussed in Section~\ref{Sec:aux:info}. In our simulations, the ensemble Burden, SKAT and MORST tests with $a_j=Beta(1,25)$ and $a_j=Beta(1,1)$ are investigated and compared with their corresponding original versions, respectively. For all the ensemble tests, the number of base tests is set to be $B=1000$. We use the prefix \textit{EN} to indicate that the test is an ensemble test.

We first examine the empirical type I error of the \textit{EN-Burden}, \textit{EN-SKAT} and \textit{EN-MORST} with $a_j = Beta(1,25)$ and $a_j=Beta(1,1)$ across a range of significance levels $\alpha = 0.05,10^{-2},10^{-3}$, $\cdots,10^{-6}$. We randomly select $10^6$ subregions from each of the 100 1Mb big regions to obtain a total $10^8$ genotype matrices. For every genotype matrix, $Y$ is then generated under the null model that $\boldsymbol{\beta}=0$ and the p-value of each ensemble test is calculated. The empirical type I error of an ensemble test is the proportion of p-values less than $\alpha$. Table~\ref{tab:typeI} presents the results of the empirical type I errors. It can be seen that the type I errors of the ensemble tests are inflated when $\alpha$ is not too small (e.g., $\alpha=0.05$) but are protected for very small $\alpha$'s (e.g., $\alpha = 10^{-4}$). This agrees with the general performance of ACAT, i.e., the Cauchy-based-approximation becomes more accurate as $\alpha$ decreases~\citep{liu2020cauchy}. In large-scale multiple testing such as WGS studies, where the significance threshold $\alpha$ is exceedingly small (e.g., $\alpha<10^{-6}$), the simulation results indicate that the p-values of \textit{EN-Burden}, \textit{EN-SKAT} and \textit{EN-MORST} can be simply calculated using the Cauchy-based-approximation~\eqref{eq:pval:approx} without correction.

\begin{table}
  \caption{\label{tab:typeI} The type I error of the ensemble tests computed over $10^{8}$ replications. Six ensemble tests are examined: \textit{EN-Burden}, \textit{EN-SKAT}, \textit{EN-MORST}, and each with $a_j = Beta(1,25)$ and $a_j = Beta(1,1)$. }
    \begin{tabular}{ccccccc}
    \toprule
    \multirow{2}[0]{*}{$\alpha$} & \multicolumn{3}{c}{\textit{Beta(1,1)}} & \multicolumn{3}{c}{\textit{Beta(1,25)}} \\
          & EN-Burden & EN-SKAT & EN-MORST & EN-Burden & EN-SKAT & EN-MORST \\ \midrule
    0.05  & 5.93$\times 10^{-2}$ & 5.96$\times 10^{-2}$ & 5.87$\times 10^{-2}$ & 6.13$\times 10^{-2}$ & 6.33$\times 10^{-2}$ & 6.10$\times 10^{-2}$ \\
    $1\times 10^{-2}$ & 1.23$\times 10^{-2}$ & 1.13$\times 10^{-2}$ & 1.15$\times 10^{-2}$ & 1.28$\times 10^{-2}$ & 1.18$\times 10^{-2}$ & 1.20$\times 10^{-2}$ \\
    $1\times 10^{-3}$ & 1.20$\times 10^{-3}$ & 1.06$\times 10^{-3}$ & 1.10$\times 10^{-3}$ & 1.26$\times 10^{-3}$ & 1.12$\times 10^{-3}$ & 1.17$\times 10^{-3}$ \\
    $1\times 10^{-4}$ & 1.14$\times 10^{-4}$ & 1.07$\times 10^{-4}$ & 1.06$\times 10^{-4}$ & 1.09$\times 10^{-4}$ & 1.00$\times 10^{-5}$ & 1.09$\times 10^{-4}$ \\
    $1\times 10^{-5}$ & 1.03$\times 10^{-5}$ & 1.02$\times 10^{-5}$ & 1.05$\times 10^{-5}$ & 1.08$\times 10^{-5}$ & 9.50$\times 10^{-6}$ & 1.05$\times 10^{-5}$ \\
    $1\times 10^{-6}$ & 9.81$\times 10^{-7}$ & 1.01$\times 10^{-6}$ & 9.75$\times 10^{-7}$ & 1.03$\times 10^{-6}$ & 1.03$\times 10^{-6}$ & 9.58$\times 10^{-7}$ \\
    \bottomrule
    \end{tabular}%
\end{table}%

We next compare the power of six ensemble tests, i.e., \textit{EN-Burden}, \textit{EN-SKAT}, \textit{EN-MORST}, and each with $a_j=Beta(1,25)$ and $a_j=Beta(1,1)$, with their corresponding original tests without ensembling, respectively. The proportion of non-zero $\beta_j$'s is set to be $20\%$ and $40\%$. For the non-zero $\beta_j$'s, two types of effect sizes are considered: (i) constant effect size, $|\beta_j| = \beta_0$ is the same for all the SNVs with non-zero effect; (ii) varied effect sizes, $|\beta_j| = \beta_0 |\log_{10} \text{MAF}_j|$ has a negative relationship with the MAFs of SNVs. The first type of effect sizes is used for comparing tests with $a_j = Beta(1,1)$, and the second type of effect sizes is used for comparing tests with $a_j = Beta(1,25)$.
The signs of the $\beta_j$'s are set to be the same in simulations for the Burden tests, or determined randomly with equal probability in simulations for SKAT and MORST. In each simulation setting, a set of equally spaced grid points of $\beta_0$'s are selected to make the power of the tests roughly range from 0.1 to 0.9. We compare the power at significance levels $\alpha = 0.05$ and $\alpha= 10^{-6}$. When $\alpha=0.05$, since the type I errors of the ensemble tests are inflated (see Table~\ref{tab:typeI}), we obtain their critical values through $10^6$ Monte Carlo simulations to make a fair comparison.

\begin{figure}[!h]
  \centering
  \includegraphics[scale=0.65]{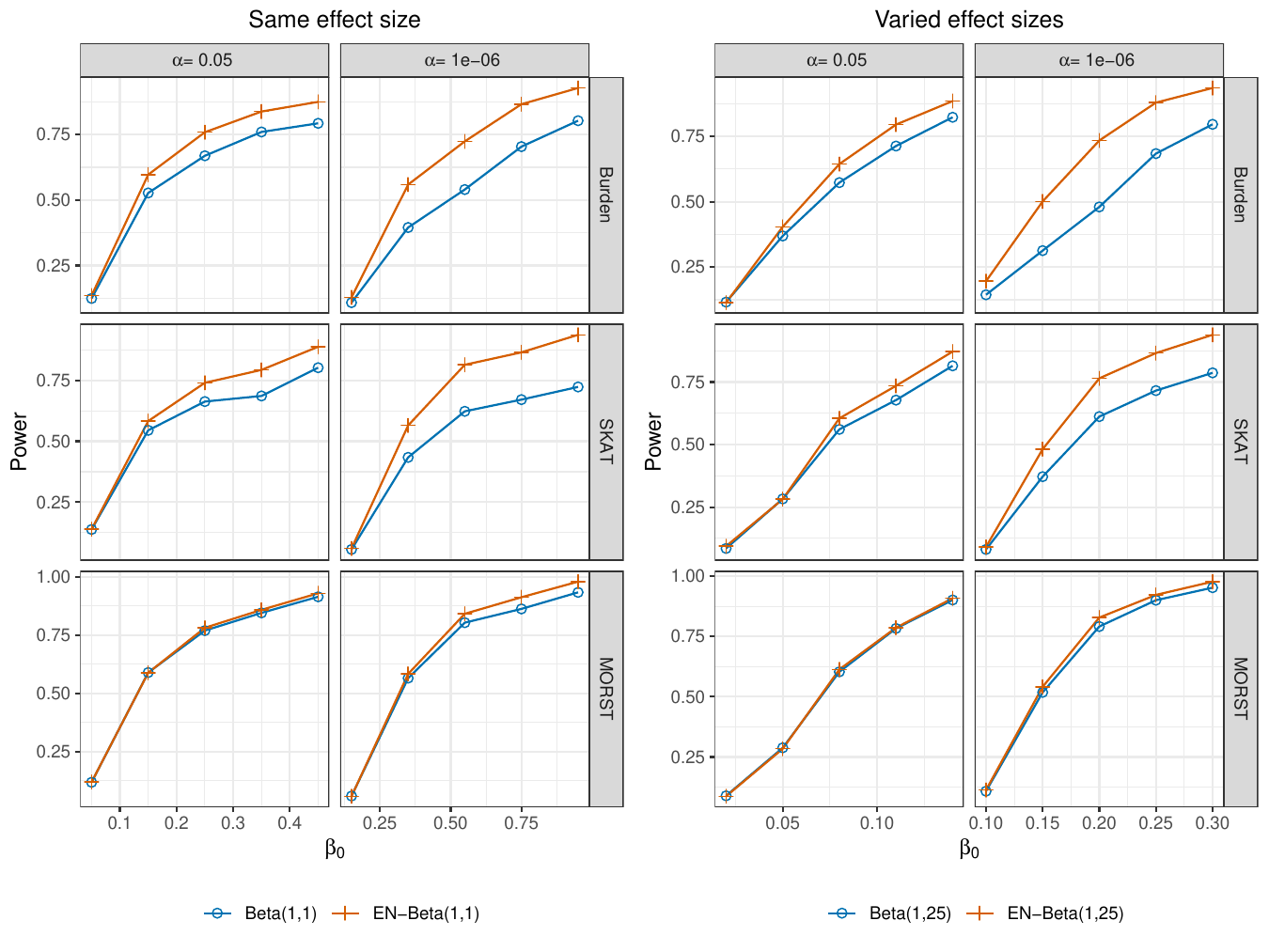}
  \caption{Power comparison of the ensemble tests with their corresponding original versions without ensembling in linear regression models with $20\%$ of $\beta$'s being non-zero. Six tests (Burden, SKAT, MORST, and each with $a_j = Beta(1,25)$ and $a_j = Beta(1,1)$) are compared with their ensemble counterparts, respectively. For the panels on the left, the effect sizes are set to be the same (i.e., $|\beta_j|=\beta_0$) and the tests with $a_j = Beta(1,1)$ are compared. For the panels on the right, the effect sizes decrease with the MAFs of SNVs (i.e., $|\beta_j| = \beta_0 |\log_{10} \text{MAF}_j|$) and the tests with $a_j = Beta(1,25)$ are compared. Two significance levels $\alpha =0.05$ and $\alpha  = 10^{-6}$ are considered. When $\alpha =0.05$, the critical values of the ensemble tests are obtained through Monte Carlo simulations. }
  \label{PowerSimuplots}
\end{figure}

Figure~\ref{PowerSimuplots} displays the power comparison results in the setting when $20\%$  $\beta_j$'s are non-zero. Each of the six ensemble tests is more powerful than or at least as powerful as the corresponding original test without ensembling. This indicates that our proposed ensemble framework can improve power for different types of tests and in situations with/without auxiliary information about the magnitude of $\beta_j$'s. We can also see that the relative power gain of the ensemble test is larger at the significance level $\alpha=10^{-6}$ than that at $\alpha=0.05$. This can be partly explained by our theoretical results as Bahadur efficiency concerns exceedingly small $\alpha$'s. Furthermore,
The power improvement of the ensemble test is more substantial for the Burden test and SKAT than that for MORST. This is because MORST already takes power robustness into account and it as a base test is more robust (or stronger) than the Burden test and SKAT. The results in the setting with $40\%$ of $\beta_j$'s being non-zero are displayed in Figure 7 in the supplementary material, and demonstrate a similar phenomenon. In addition, we also provide the simulation results about the power comparisons of the three ensemble tests (i.e., EN-Burden, EN-SKAT and EN-MORST) under different alternatives in Section 6 in the supplementary material. The results are similar to the comparisons of the three original tests~\citep{liu2019acat,liu2020minimax}.

\section{Analysis of the ARIC whole-genome sequencing data}\label{Sec:realdata}

We apply the proposed ensemble Burden, SKAT and MORST tests to the analysis of WGS data from the ARIC study~\citep{morrison2017practical}.  As with the simulations in Section~\ref{Sec:simu}, the tests with $a_j = Beta(1,1)$ and $a_j = Beta(1,25)$ are applied and we use a prefix \textit{EN} to denote the ensemble tests, e.g., \textit{EN-Burden}.

Following the sample-level quality control detailed in~\cite{morrison2017practical}, we have around 55 million SNVs in 1860 African Americans (AAs). Our analysis focuses on SNVs with MAF less than 0.05, and aims to identify SNV-sets associated with lipoprotein(a) (Lp(a)) and neutrophil count in AAs. Both Lp(a) and neutrophil count are quantitative traits and risk factors of cardiovascular diseases. The SNV-sets are defined according to a sliding window approach~\citep{morrison2017practical}, which chooses physical windows of 4000 base pairs (4kb) in length, starts at 0 base pair for each chromosome, and uses a skip length of 2kb. This results in a total of 1,337,673 SNV-sets (i.e., 4kb windows). The distribution of the
number of SNVs in a sliding window has a median of 60 and is heavily skewed to the right. The Bonferroni correction is used to adjust for multiple testing, and the genome-wide significance threshold is set to be $\alpha = 3.75 \times 10^{-8}$.
Because both Lp(a) and neutrophil count have skewed distributions, rank-based inverse normal transformation is performed and the transformed phenotypes are used as the outcomes in the analysis. We adjust for age, sex and the first three of principal components about population stratification~\citep{price2006principal} for both phenotypes, and additionally smoking status for neutrophil count.

\begin{figure}[!h]
  \centering
  \includegraphics[scale=0.75]{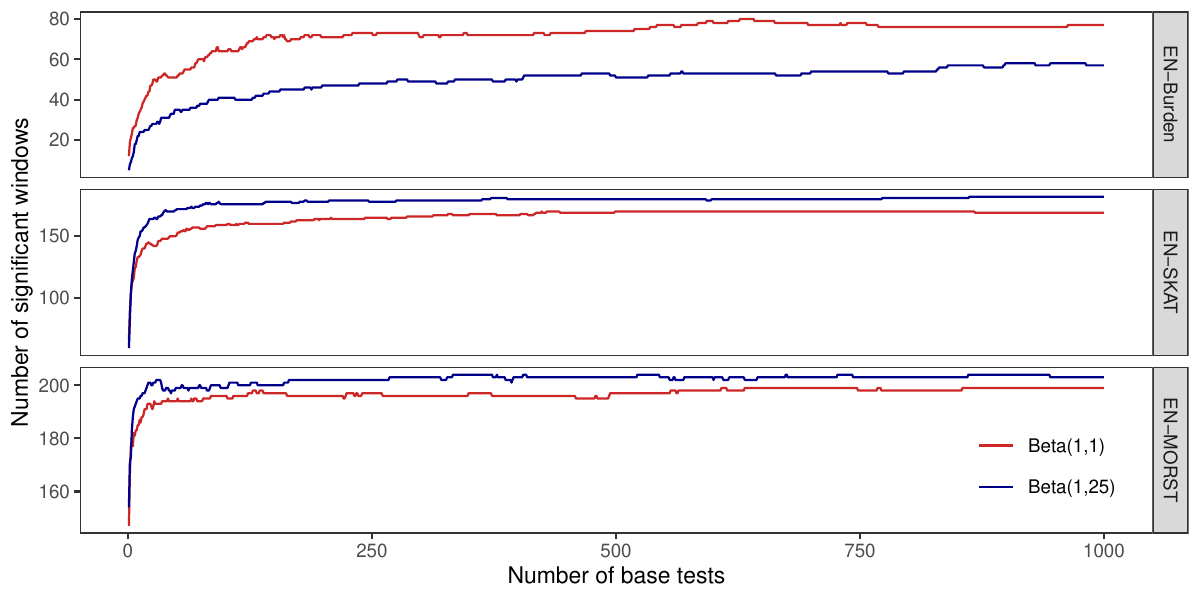}
  \caption{The path of the numbers of significant sliding window identified by each ensemble test as the number of base tests changes in the rare variant association analysis of lipoprotein(a) using the ARIC Whole Genome Sequencing Study of African American. The $x$-axis is the number of base tests used in the ensemble test. The panels from top to bottom correspond to the results for the ensemble Burden test, ensemble SKAT, and ensemble MORST, respectively. In each panel, the two lines corresponds to the ensemble test using  $a_j = Beta(1,1)$ and $a_j = Beta(1,25)$, respectively. The significance threshold $\alpha$ is $3.75\times 10^{-8}$.  }
  \label{Fig:SigWinPath}
\end{figure}

Figure~\ref{Fig:SigWinPath} shows how the numbers of significant sliding windows changes as $B$ (i.e., the number of base tests) increases for each of the six ensemble tests in the analysis of Lp(a). For all the ensemble tests,
 the number of significant windows almost stabilizes at $B=200$ base tests and increases slowly as $B$ continues to go up. This indicates that the power of the ensemble tests becomes stable when the number of base tests are sufficiently large. This phenomenon is analogous to that in random forests where the prediction accuracy of a random forest would stabilize as the number of random trees increases. In addition, we can see that \textit{EN-SKAT} and \textit{EN-MORST} stabilize faster than \textit{EN-Burden}. This is expected because the base test of \textit{EN-SKAT} or \textit{EN-MORST} is more robust than that of \textit{EN-Burden}, i.e., the power of the base test of \textit{EN-SKAT} or \textit{EN-MORST} is less variable than that of \textit{EN-Burden}. Hence, less base tests are required by \textit{EN-SKAT} or \textit{EN-MORST} to stabilize power.
  The results for neutrophil count are similar and provided in Figure 8 in the supplementary materials. Hence, we set the number of base test $B=1000$, where the results of our proposed ensemble tests are already quite stable.

Table~\ref{tab:sig:windows} shows the number of significant sliding windows detected by the original Burden, SKAT and MORST tests and their ensemble counterparts. For the class of Burden tests, the ensemble tests identify much more significant 4kb windows compared to their respective original counterpart tests with $a_j = Beta(1,1)$, where no information about effect sizes is used, and $a_j=Beta(1,25)$, where the MAF-based information is incorporated. The superior performance of the proposed ensemble Burden tests is   consistent across the analyses of different phenotypes. For the class of SKAT tests, substantial and consistent power improvement is also achieved by the corresponding ensemble tests. The MORST test already takes power robustness into account and is robust across different signal strengthes~\citep{liu2020minimax}. In other words, MORST is not a very weak base test to start with. But we still observe power gain of the ensemble MORST tests, although the power gain is smaller compared to that for the Burden and SKAT tests. In summary, the real-data analysis results also demonstrate that the proposed ensemble tests can greatly enhance the power and are at least as powerful as the original tests.

\begin{table}
  \caption{\label{tab:sig:windows} The number of significant 4kb sliding windows identified by each test that are associated with
lipoprotein(a) or neutrophil count in the analysis of the ARIC Afrian American WGS data. The original and ensemble  Burden, SKAT and MORST tests using the prior weights $a_j=Beta(1,1, MAF_j$) and $a_j=Beta(1,25, MAF_j)$ are compared, respectively.
 For all the ensemble tests (denoted with a prefix EN), the number of base tests is set to be $B=1000$. The significance threshold $\alpha$ is $3.75\times 10^{-8}$ based on the Bonferroni correction.}
 \centering
    \begin{tabular}{ccccc}
    \toprule
    \multicolumn{5}{c}{\textbf{Burden}} \\
          & Beta(1,1) & EN-Beta(1,1) & Beta(1,25) & EN-Beta(1,25) \\ \midrule
    Lipoprotein(a) & 8     & 79    & 6     & 54 \\
    Neutrophil count & 20    & 79    & 8     & 43 \\\toprule
    \multicolumn{5}{c}{\textbf{SKAT}} \\
          & Beta(1,1) & EN-Beta(1,1) & Beta(1,25) & EN-Beta(1,25) \\\midrule
    Lipoprotein(a) & 103   & 171   & 127   & 184 \\
    Neutrophil count & 113   & 149   & 62    & 126 \\\toprule
    \multicolumn{5}{c}{\textbf{MORST}} \\
          & Beta(1,1) & EN-Beta(1,1) & Beta(1,25) & EN-Beta(1,25) \\\midrule
    Lipoprotein(a) & 188   & 196   & 199   & 204 \\
    Neutrophil count & 184   & 193   & 134   & 160 \\\bottomrule
    \end{tabular}%
\end{table}%

\section{Discussion}\label{Sec:diss}

In this paper, we proposed a flexible and computationally efficient ensemble testing framework to develop powerful global tests for a certain class of alternatives.  The proposed framework was then used to construct specific ensemble tests for four different classes of alternatives.
The key component of the ensemble framework is to use certain random procedure (i.e., $\Theta_i$) to deal with or characterize the features of the class of the alternatives, and the improvement of the ensemble strategy for testing can be similarly explained as that for random forests via certain kinds of variability reduction (Section~\ref{Secsub:explanation}). The real data analysis and simulation studies demonstrate that the ensemble tests are at least as powerful as and could be much more powerful than the existing tests.

It should be noted that all the proposed ensemble tests are still valid with protected type I errors even when the true alternative does not belong a particular class under which a test is constructed, as their global null hypothesis is the same. However, they will be subject to power loss. For example, the ensemble Burden test has the correct type I error rate when the effect signs are different and is still valid. But it has a lower power compared to the ensemble SKAT and MORST tests under such alternatives.

The theoretical properties of our proposed ensemble tests are established with respect to Bahadur efficiency in their respective settings. As mentioned in Remark~\ref{Rmk1}, the Bahadur efficiency results in this work are about specific classes of alternatives and cannot be directly generalized to general tests using ACAT. It is of future interest to study the Bahadur efficiency about ACAT itself.
The Bahadur efficiency only reflects the testing power in the limiting situation when the significance level $\alpha$ goes to 0, and may partly explain the superior performance of the ensemble tests in our application of SNV-set tests in analysis of WGS data. It is of great future interest to investigate the theoretical properties of the ensemble tests from other perspectives, e.g., the power robustness of ensemble tests, which could be helpful for guiding the development of other ensemble methods.

 Motivated by WGS association studies, this work focuses on the large-scale multiple testing setting where the dimension $p$ in a test is often not large. It is also interesting to extend the ensemble tests to the high-dimensional setting. When $p$ is large, the number of base tests $B$ required by the ensemble tests could increase, as indicated by theorems~\ref{Thm:burden},~\ref{Thm:SKAT} and~\ref{Thm:ENchisq}. Hence, better sampling strategies and/or base test designs are needed. One way is to adapt a warm-start idea and use the data-splitting procedure. Specifically, we divide $n$ i.i.d. samples into two datasets, i.e., one small dataset with the sample size $n_s = o(n)$  and a large dataset with the sample size $n_l = n - n_s$. The small dataset is first used to obtain some preliminary information about the parameters of interest. Testing is then performed on the large dataset, with the preliminary information from the small dataset being incorporated to narrow down the space where the random components $\Theta_i$'s are sampled. For instance, in the sparse testing problem considered in Section~\ref{Subsec:sparse:testing}, we can first use the small dataset to rule out some components of $\mathbf{Z}$ that are very unlikely to have a non-zero mean. In other words, we obtain a subset of $\{1,2,\cdots, p\}$ that corresponds to the indices of the retained components. Then, when applying the ensemble subset chi-squared test to the large dataset, we draw $m$ indices from this subset. In this way, the number of base tests $B$ could be reduced and theoretically, $B$ can be reduced to $O(1)$ to achieve the Bahadur efficiency. More details about the theory and some simulation results of this data-splitting procedure are provided in Section 7 in the supplementary materials.

While we only study the problem of testing a global null in this work, the proposed ensemble framework is general and can also be applied to other testing problems such as testing a composite null. One future research direction is to explore whether the ensemble strategy could be helpful for testing in general. In addition,
the ensemble idea is very intuitive and has been successfully applied to develop a variety of effective learning methods, such as random forest and boosting.
Random forest uses parallel ensemble while boosting uses sequential ensemble.
Our proposed ensemble testing framework adapts the parallel ensemble idea in the same spirit of random forests for regression. Developing other ensemble methods for testing, such as sequential ensemble methods, could be an interesting topic for future research.

\section*{Acknowledgements}

X. Lin's Research was supported by  the US National Genome Research Institute grants U01-HG009088 and U01-HG012064,  the US National Institute of Heart, Lung and Blood R01-HL113338, and the US National Cancer Institute R35-CA197449 and U19-CA203654. Y. Liu's research is partially supported by the National Natural Science Foundation of China (Nos. 12001443 and 12371282).
The authors thank the Associate Editor and the referees for very helpful and constructive comments, which have greatly helped improve the paper.

\bibliographystyle{chicago}
\bibliography{ET}

\end{document}